\documentclass[preprints,article,accept,moreauthors,pdftex]{mdpi} 

\firstpage{1} 
\pubvolume{xx}
\issuenum{1}
\articlenumber{5}
\pubyear{2019}
\copyrightyear{2019}
\history{Received: date; Accepted: date; Published: date}

\usepackage{amsmath}
\usepackage{amssymb}
\usepackage{graphicx}
\usepackage{dcolumn}
\usepackage{bm}
\usepackage{latexsym}
\usepackage{color}
\usepackage{xcolor}

\Title{Physics and phenomenology of weakly magnetized, relativistic astrophysical shock waves}

\Author{Arno Vanthieghem $^{1,2,3,*}$
, Martin Lemoine $^{2}$
, Illya Plotnikov $^{4}$
, Anna Grassi $^{1,5}$
, Mickael Grech $^{5}$
, Laurent Gremillet $^{6}$
and Guy Pelletier $^{7}$
}

\AuthorNames{Arno Vanthieghem, Martin Lemoine, Anna Grassi, Mickael Grech, Laurent Gremillet, Illya Plotnikov and Guy Pelletier}

\address{%
$^{1}$ High Energy Density Science Division, SLAC National Accelerator Laboratory, Menlo Park, California 94025, USA\\
$^{2}$ Institut d'Astrophysique de Paris, CNRS -- Sorbonne Universit\'e, 98 bis boulevard Arago, F-75014 Paris\\
$^{3}$  Sorbonne Universit\'e, Institut Lagrange de Paris (ILP), 98 bis bvd Arago, F-75014 Paris, France\\ $^{4}$ Institut  de  Recherche  en  Astrophysique  et  Plan\'etologie,  Universit\'e de  Toulouse III, OMP, CNRS, 9  avenue  Colonel  Roche,  BP  44346 - 31028,  Toulouse,  France\\
$^{5}$  LULI, CNRS, CEA, Sorbonne Universit\'e, Ecole Polytechnique, Institut Polytechnique de Paris, F-91128 Palaiseau Cedex, France \\
$^{6}$ CEA, DAM, DIF, F-91297 Arpajon, France\\
$^{7}$ UJF-Grenoble, CNRS-INSU, Institut de Planétologie et d’Astrophysique de Grenoble (IPAG), F-38041 Grenoble, France
}

\corres{Correspondence: vanthieg@slac.stanford.edu}

\abstract{Weakly magnetized, relativistic collisionless shock waves are not only the natural offsprings of relativistic jets in high-energy astrophysical sources, they are also associated with some of the most outstanding displays of energy dissipation through particle acceleration and radiation. Perhaps their most peculiar and exciting feature is that the magnetized turbulence that sustains the acceleration process, and (possibly) the secondary radiation itself, is self-excited by the accelerated particles themselves, so that the phenomenology of these shock waves hinges strongly on the microphysics of the shock. In this review, we draw a status report of this microphysics, benchmarking analytical arguments with particle-in-cell simulations, and extract consequences of direct interest to the phenomenology, regarding in particular the so-called microphysical parameters used in phenomenological studies.
}


\begin{document}

\section{Introduction}\label{sec:intr}

The generation of fast and powerful outflows appears to be a common trait of all high-energy astrophysical sources, often accompanied by outstanding dissipative phenomena observed all across the electromagnetic spectrum. The bright, persistent or sporadic multiwavelength nonthermal radiation emitted by the jets of radio-galaxies, blazars and microquasars,  {\it e.g.}~\cite{Romero_2017,Rieger_2018}, the prompt and long-term afterglow emission of gamma-ray bursts from low to high energies, {\it e.g.}~\cite{Piran_2004,Kumar_2015a,Naval_2018}, the shining pulsar wind nebulae, {\it e.g.}~\cite{Bykov_2017,Amato_2020}, the electromagnetic counterparts of recent gravitational wave events, {\it e.g.}~\cite{Abbott_2017b}, or even the very generation of cosmic rays and neutrinos at extreme energies, {\it e.g.}~\cite{Halzen_2017,Meszaros_2019}... all those represent quite remarkable examples.

Although particle energization can take place in a variety of environments and through diverse acceleration mechanisms, the dissipation agent at play is often a collisionless shock front. It converts a substantial fraction of a (low entropy) ordered form of energy into a (high entropy) gas of suprathermal particles that then radiate through their interactions with ambient fields ({\it e.g.}, via synchrotron, inverse Compton scattering, Bremsstrahlung, etc.).  This ordered form of energy may be  kinetic and/or electromagnetic in nature. It is generally sourced by a central object, which consumes its rotational or magnetic energy reservoir, or which draws hydromagnetic energy from its surroundings.

Those shock waves are said to be collisionless because the mean free path for (Coulomb) binary particle interactions far exceeds the relevant length scales, in particular the thickness of the shock, and this very property makes them ideal sites of particle acceleration, {\it e.g.}~\cite{Blandford_1987}, proceeding in rather extreme regimes of relativistic plasma astrophysics~\cite{Kirk_1999,Bykov_2012,Sironi_2015}. In the very high-energy Universe, the outflows can be so powerful that these shocks move at relativistic velocities $\beta_{\rm sh} \equiv v_{\rm sh}/c \sim 1$ into the surrounding plasma. To characterize these phenomena, it is more convenient to speak in terms of four- rather than three-velocity, defining in particular the four-velocity (in units of $c$) of the shock in the unshocked plasma frame,  $u_{\rm sh}=\beta_{\rm sh}/\sqrt{1-\beta_{\rm sh}^2}$, its corresponding Lorentz factor $\gamma_{\rm sh}=u_{\rm sh}/\beta_{\rm sh}$, and to introduce the ambient magnetization, $\sigma=u_{\rm A}^2$, with $u_{\rm A}=B/\sqrt{4\pi n m c^2}$ the Alfv\'en four-velocity (units of $c$) of the ambient plasma, which depends on the magnetic field strength $B$, the plasma proper number density $n$ and particle mass $m$.

These two parameters, $u_{\rm sh}$ and $\sigma$, together with the composition of the ambient plasma through which the shock is propagating, control the physics of the shock, hence the fate of dissipation. The termination shock of pulsar winds, which separates the inner (unshocked) wind region from the pulsar wind nebula, likely represents the most extreme regime, with $u_{\rm sh}\gtrsim 100-1000$, possibly much larger, and $\sigma\sim 0.1$, although this magnetization could be significantly lower in the narrow equatorial region~\cite{Kirk09}. By contrast, the external ultrarelativistic shock that precedes the ejecta of gamma-ray bursts, interacts with a medium of weak magnetization, $\sigma\sim10^{-9}$ (for the interstellar medium), possibly as large as $10^{-5}$ in magnetized circumstellar winds, while $u_{\rm sh}$ decreases with time from values as large as $100-1000$, down to the subrelativistic regime~\cite{Piran_2004}. Inside gamma-ray burst outflows, radio-galaxy and blazar outflows, or even microquasars, internal shocks can occur between layers of material propagating at different four-velocities, triggering mildly relativistic [meaning $u_{\rm sh}\sim\mathcal O(1)$] shock waves, {\it e.g.,}~\cite{Rees_1978, Daigne_1998, Malzac_2013, Baring_2017}. In such objects, the magnetization is generally unknown, and may take arbitrarily large or small values. The termination shock of AGN jets and the reverse shock propagating back into the gamma-ray burst ejecta are also believed to be mildly relativistic. There as well, the magnetization is generally regarded as a parameter.

The landscape of relativistic, collisionless shock waves is thus quite broad, and much of it remains to be uncovered. This review will deal with the regime of weak magnetization, meaning $\sigma\lesssim 10^{-3}$. As the above examples suggest, this regime is likely applicable to many types of high-energy astrophysical sources. It is also a regime in which particle acceleration is known to be particularly efficient, and in which the turbulence that sustains the Fermi process is generated through plasma microinstabilities by the accelerated particles themselves. Under such conditions, one cannot address the physics of particle acceleration, and its phenomenological signatures in high-energy and multi-messenger astrophysics, without delving into the plasma physics of the shock and of its accompanying microturbulence.

The outline of this review is as follows. In Sec.~\ref{sec:struc}, we first sketch the general structure of a shock front on the (fluid) scales on which it appears as a discontinuity, then zoom in on `microphysical' scales in Sec.~\ref{sec:micr}, so as to explain how particles build the effective magnetized microturbulence that promotes their acceleration. We then extract phenomenological consequences from this microphysics in Sec.~\ref{sec:phen}, and discuss their potential impact in high-energy astrophysics, to conclude with a summary in Sec.~\ref{sec:summ}. Throughout, we use units in which $k_{\rm B}=c=1$, and consider a metric signature $(-,+,+,+)$. Quantities evaluated in a particular reference frame, say $\mathcal R_{\rm d}$ (with d for downstream), are annotated with the corresponding subscript $_{\vert\rm d}$. Velocities written as $u$ are understood as four-velocities, while $\beta$ denotes a three-velocity. All densities and temperatures are expressed in the proper frame.

\section{The hydrodynamical view}\label{sec:struc}

In a purely hydrodynamical picture, shock waves form through the interpenetration of fluids at a relative velocity $u_{\rm rel}$ larger than their sound velocity $u_{\rm so}$. In the astrophysical context, the relevant wave velocity becomes the fast magnetosonic velocity $u_{\rm F} \simeq {\rm max}\left[u_{\rm so}, u_{\rm A} \right]$.

\subsection{Shock velocity}

A general situation is one in which an ejecta with proper density $n_{\rm ej}$, bulk velocity  $u_{\rm ej\vert ext}$ penetrates a medium at rest with proper density $n_{\rm ext}$. Here, we assume a relativistic interaction, that is, $\lvert u_{\rm ej\vert ext} \rvert \gg1$. This generally gives rise to a double-shock structure: one (forward) propagating through the external medium, and one (reverse) propagating through the ejecta. From left (ejecta) to right (external medium), the overall fluid structure is then composed of four zones: the unshocked ejecta, the shocked ejecta, the shocked external medium and finally the unshocked external medium. The reverse shock represents the transition layer between the unshocked and shocked ejecta, while the forward shock corresponds to that between the unshocked and shocked external media. Finally, the shocked ejecta and shocked external medium are in pressure equilibrium, separated by a contact discontinuity surface. Both shocked fluids, on either side of the contact discontinuity, move at the same velocity.

To derive the characteristics of these shock fronts, it is best to study the problem in the blast (\emph{i.e.}, the shocked medium) reference frame $\mathcal R_{\rm b}$ in which the two unshocked plasmas carry equal momentum flux densities. This reference frame is that in which the contact discontinuity lies at rest. Assuming for simplicity that the momentum flux is dominated by the kinetic ram pressure of each fluid (hence neglecting the internal and electromagnetic energies), we have
\begin{equation}
    u_{\rm ej\vert b}^2 n_{\rm ej} = u_{\rm ext\vert b}^2 n_{\rm ext}\,.
    \label{eq:bframe}
\end{equation}
One may then consider three typical situations and prove the following, see {\it e.g.}~\cite{1995ApJ...455L.143S}:
\begin{enumerate}
    \item Assume first that the blast frame is close to the comoving reference frame of the ejecta, which means that the forward shock is ultrarelativistic, while the reverse shock propagating is subrelativistic. This means $\beta_{\rm ej\vert b}<1$ and $u_{\rm ext\vert b}^2 \simeq u_{\rm ej\vert ext}^2 \gg 1$. From Eq.~\eqref{eq:bframe} above, one then infers  $\beta_{\rm ej\vert b}\simeq u_{\rm ej\vert ext} \left(n_{\rm ext}/n_{\rm ej}\right)^{1/2}$. This situation occurs when $u_{\rm ej\vert ext}  \ll (n_{\rm ej}/n_{\rm ext})^{1/2}$ (since $\beta_{\rm ej\vert b}<1$) and $n_{\rm ext} \ll n_{\rm ej}$. \\
    
     \item Conversely, the blast frame may be close to the external medium frame, in which case the forward shock becomes subrelativistic while the reverse shock is ultrarelativistic.  Using arguments similar to those detailed above, one then finds $\beta_{\rm b\vert ext}\simeq u_{\rm ej\vert ext} (n_{\rm ej}/n_{\rm ext})^{1/2}$, and $u_{\rm ext\vert b}^2<1$. This limit applies when $n_{\rm ej} \ll n_{\rm ext}$ and $u_{\rm ej\vert ext} \ll (n_{\rm ext}/n_{\rm ej})^{1/2}$.\\
     
    \item In between those two extreme limits, both forward and reverse shocks are truly relativistic. Under these conditions, the shock velocities satisfy the following hierarchy: $1\,\ll\,u_{\rm ej\vert b}\,,\,\, u_{\rm b\vert ext}\,\ll\,u_{\rm ej\vert ext}$. It then becomes convenient to approximate the relative velocity $u_{\rm ej\vert b}$ as  $u_{\rm ej\vert b}\simeq (u_{\rm ej\vert ext}/u_{\rm b\vert ext}-u_{\rm b\vert ext}/u_{\rm ej\vert ext})/2 \simeq u_{\rm ej\vert ext}/(2 u_{\rm b\vert ext})$. Inserting in Eq.~(\ref{eq:bframe}) then gives $u_{\rm b\vert ext} \simeq u_{\rm ej\vert ext}^{1/2}\left(n_{\rm ej}/4n_{\rm ext}\right)^{1/4}$. Similarly, one also obtains $u_{\rm ej\vert b} \simeq u_{\rm ej\vert ext}^{1/2}\left(n_{\rm ext}/4n_{\rm ej}\right)^{1/4}$.
    
\end{enumerate}

\begin{figure}
    \centering
	\includegraphics[width=0.8\textwidth]{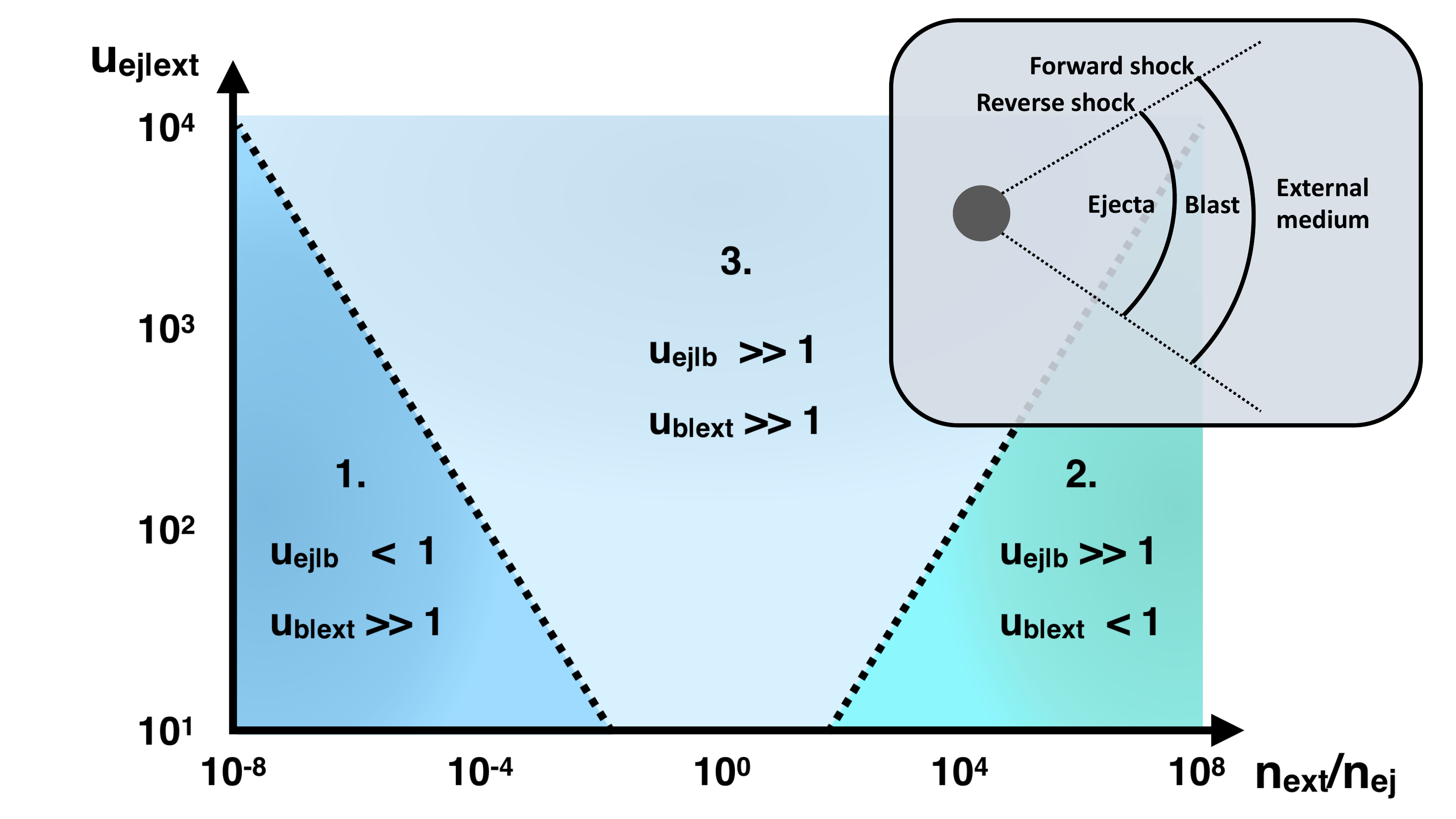}
	\caption{
	\label{fig:blast} Phase diagram of relativistic ejecta $(| u_{\rm ej|ext} | \geq 10)$ penetrating the external medium in the plane $\left( n_{\rm ext}/n_{\rm ej},\, | u_{\rm ej | ext} | \right)$. The two dotted diagonals delimit the transition between the three typical regimes for the forward and reverse shocks in the blast wave frame: (1.) a subrelativistic reverse shock and an ultrarelativistic forward shock; (2.) an ultrarelativistic reverse shock and a subrelativistic forward shock; (3.) relativistic reverse and forward shocks. The inset illustrates the general configuration for a relativistic jet.}
\end{figure}

Those different configurations are illustrated in the phase diagram of Fig.~\ref{fig:blast}. Both situations 1 and 3 above may govern the physics of the interaction between the ejecta of a gamma-ray burst and the circumstellar medium. Situation 2 is more typical of a pulsar wind nebula, where the termination (reverse) shock is ultrarelativistic while the forward shock propagating in the supernova remnant is well subrelativistic. In this case, however, the electromagnetic nature and the three-dimensional geometry of the wind, which implies dilution through expansion, slightly modify the above arguments. In particular, the wind ram pressure is given by $L_{\rm w}/(4\pi r^2 c)$ in terms of the wind luminosity $L_{\rm w}$ and radial distance $r$ from the source. Of course, were the velocity of the shock to become smaller than the effective magnetosonic velocity of the ambient medium, it would turn into a compressional wave, and particle acceleration would be quenched. In the following, we study the dynamics and structure of a relativistic shock, which could thus represent either the forward of reverse shock, depending on the ambient conditions, as described above.

\subsection{Shock jump conditions}

Let us now narrow the picture on one relativistic shock front, assuming that the relative velocity $\beta_{\rm u\vert d}$ between the unshocked (u) and shocked (d) plasmas has been obtained using the above arguments, {\it viz.} $\beta_{\rm u\vert d}=\beta_{\rm ext \vert b}$ for the forward shock, $\beta_{\rm u\vert d}=\beta_{\rm ej\vert b}$ for the reverse shock. Following standard conventions, u (resp. d) refers to the upstream (resp. downstream) plasma, since, when viewed in the reference frame in which the shock lies at rest ($\mathcal{R}_{\rm s}$), the unshocked plasma is inflowing (from upstream), becomes compressed and heated through the shock, and outflows (into the downstream). The relativistic jump conditions can be obtained from the conservation equations of four-current and energy-momentum:
\begin{align}
    \partial_\nu J^\nu &= 0 \,, &
    \partial_\nu T^{\mu \nu} &= 0 \,,
\end{align}
where $J^{\nu} = \left( \gamma n, \boldsymbol{u} n \right)$ represents the four-current of the plasma ($\gamma=u^0$ the Lorentz factor), and $T^{\mu \nu}$ the energy-momentum tensor. For an ideal fluid, $T^{\mu \nu} = w u^\mu u^\nu + p \eta^{\mu \nu}$, with $w = p+e$ the enthalpy density, $p$ the pressure and $e$ the energy density. Integration of these conservation equations across the shock surface gives the shock crossing conditions, which are well-known in the hydrodynamic or weakly magnetized limit, of interest to this review~\cite{Taub_1948,Akhiezer_1959,Lichnerowicz_1970,Blandford_1976,Kirk_1999,Peer_2017}. Here, we derive these conditions in a much simpler way, namely, by integrating the conservation equations in the downstream rest frame directly, rather than in the shock frame, following~\cite{Lemoine_2016}. To this effect, we use the fact that in any frame, the shock crossing conditions can be written~\cite{Kirk_1999}
\begin{align}
    \left[ J^{\nu} \right] l_\nu &= 0 \,, &
    \left[ T^{\mu \nu} \right] l_\nu &= 0 \,,
\end{align}
where $l_\nu$ represents the four-vector that is normal to the shock surface. In the downstream rest frame,  $l_\nu = \left( -\gamma_{\rm s|d} \beta_{\rm s|d} , \gamma_{\rm s|d}, 0, 0  \right)$, hence we obtain
\begin{align}
    &\gamma_{\rm u|d} n_u \left( \beta_{\rm u|d} - \beta_{\rm s|d} \right) = - n_{\rm d} \beta_{\rm s|d}\,, \\
    &\gamma^2_{\rm u|d} w_{\rm u} \left( \beta_{\rm u|d} - \beta_{\rm s|d} \right) + \beta_{\rm s|d} p_{\rm u } = - \beta_{\rm s|d} \left( w_{\rm d} - p_{\rm d}\right) \,, \\
    &\gamma^2_{\rm u|d} \beta_{\rm u|d} w_{\rm u} \left( \beta_{\rm u|d} - \beta_{\rm s|d} \right) +  p_{\rm u } =  p_{\rm d} \,.
\end{align}
In the case of a strong shock ($p_{\rm u}\ll w_{\rm u}$, or, equivalently, $T_{\rm u} \ll m$), the shock jump conditions reduce to
\begin{align}
    \frac{n_{\rm d}}{n_{\rm u}} &= \gamma_{\rm u|d} \left( 1- \frac{\beta_{\rm u|d}}{\beta_{\rm s |d}} \right) \,,&
    \frac{p_{\rm d}}{n_{\rm u}} &= \gamma_{\rm u|d}^2 \beta_{\rm u|d} \left( \beta_{\rm u|d} - \beta_{\rm s|d} \right) \,, &
    \frac{T_{\rm d}}{m} &= - \gamma_{\rm u|d} \beta_{\rm s|d} \beta_{\rm u|d} \,,\\
    \beta_{\rm s|d} &= - \frac{\hat\Gamma_{\rm d} - 1}{\beta_{\rm u|d}} \,,&
    \gamma_{\rm u\vert s} &= \gamma_{\rm u\vert d}\gamma_{\rm s\vert d}\left(1-\beta_{\rm u\vert d}\beta_{\rm s\vert d}\right)\,,& &
\end{align}
where $\hat\Gamma_{\rm d}$ is the adiabatic index of the downstream plasma. For an ultrarelativistic shock propagating towards $+x$, one has $\beta_{\rm u\vert d} \simeq -1$. If $\hat\Gamma_{\rm d}=4/3$, as is relevant to a relativistically hot plasma in 3D, the shock speed relative to the ambient medium is $\beta_{\rm s|d}\simeq +1/3$, corresponding to a Lorentz factor $\gamma_{\rm sh}\equiv \gamma_{\rm u\vert s}\simeq\sqrt{2}\gamma_{\rm u\vert d}$. Furthermore, $T_{\rm d}\simeq \gamma_{\rm sh} m/(3\sqrt{2})$, and the ratio between proper densities is $n_{\rm d}/n_{\rm u}=4 \gamma_{\rm u\vert d}$. The compression ratio is defined as the ratio of apparent densities in the shock front frame, which, by virtue of current conservation, is given by $\gamma_{\rm s\vert d}n_{\rm d}/(\gamma_{\rm u\vert s}n_{\rm u})= \beta_{\rm u\vert s}/\beta_{\rm s\vert d}=3$.

Particle-in-cell (PIC) numerical simulations are often restricted to 2D3V (meaning 2D in configuration space and 3D in momentum space) due to limited computational resources. In the unmagnetized case, the relevant adiabatic index is then that of a 2D relativistic gas, $\hat\Gamma_{\rm d}=3/2$, which leads to a shock speed $\beta_{\rm s|d}\simeq+1/2$, a shock Lorentz factor $\gamma_{\rm sh}=\gamma_{\rm u\vert s}\simeq\sqrt{3}\gamma_{\rm u\vert d}$, the ratio between proper densities $n_{\rm d}/n_{\rm u}=3 \gamma_{\rm u\vert d}$ \footnote{Let us note that for a simulation performed in the downstream rest frame, the apparent density of the upstream flow is equal to $n_{\rm u, app} = n_{\rm u} \gamma_{\rm u\vert d}$. This results in an apparent density of the downstream plasma equal to $4 n_{\rm u, app}$ in 3D simulations and to $3 n_{\rm u, app}$ in 2D PIC simulations (see, black line in Figure \ref{fig:profs})}, and a compression ratio $\gamma_{\rm s\vert d}n_{\rm d}/(\gamma_{\rm u\vert s}n_{\rm u})= \beta_{\rm u\vert s}/\beta_{\rm s\vert d}=2$.

\subsection{Relativistic Fermi acceleration}

Particles can gain energy at a relativistic shock front through repeated bounces on the magnetized plasmas up- and downstream of the shock, much as in the well-known subrelativistic first-order Fermi process~\cite{Fermi_1949, Drury_1983, Blandford_1987}. The particle acceleration mechanisms have been reviewed extensively elsewhere~\cite{Ellison_1990, Bednarz_1996, Pelletier_2001, Lemoine_2006, Niemiec_2006, Bykov_2012, Sironi_2015, Marcowith_2020}, so we will simply stress some important features of the relativistic regime:
\begin{enumerate}
    \item Given that $\beta_p \sim \beta_{\rm sh}$ and $\beta_{\rm sh}\simeq 1$, with $\beta_p$ a particle velocity, accelerated particles do not diffuse spatially in the upstream plasma before returning to the shock front. They rather undergo small angle diffusion through an electromagnetic microturbulence, or small angle deflection in a large-scale magnetic field, until their parallel velocity along the shock normal ({\it i.e.}, the direction of propagation of the shock front), becomes smaller than $\beta_{\rm sh}$. At that point, the shock catches up with the particle, and the latter is thus sent downstream~\cite{Gallant_1999, Achterberg_2001}.\\
    
    \item As a consequence, if the particle spends a time $t_{\rm res\vert u}$ in the upstream, the distance between the shock front and the particle is of the order of  $\ell\simeq (\beta_p-\beta_{\rm sh})\,t_{\rm res\vert u}\simeq t_{\rm res\vert u}/\left(2\gamma_{\rm sh}^2\right)$. The corresponding region, located  immediately upstream of the shock front, where the accelerated particles mix with the unshocked plasma in the course of their Fermi cycles across the shock surface, is called the \emph{shock precursor}. It is therefore of very limited extent in the relativistic regime~\cite{Milosavljevic_2006b}. This has important implications for the development of plasma instabilities, because only those whose growth length scale is short enough, can be excited on the precursor crossing timescale~\cite{Pelletier_2008, Lemoine_2010, Lemoine_2014b}.\\
    
    \item While the spectral index of the momentum spectrum of the accelerated population scales with the shock three-velocity in the sub-relativistic regime, it reaches an asymptotic value $s\simeq 2.2$ in the relativistic regime $u_{\rm sh}\gg 1$ \citep{Bednarz_1998, Gallant_1999, Kirk_2000, Achterberg_2001, Ellison_2002, Lemoine_2003, Niemiec_2004, Keshet_2005, Warren_2015}, see also~\cite{Spitkovsky_2008a, Nishikawa_2009, Keshet_2009, Martins_2009, Haugbolle_2011, Sironi_2013} for {\it ab initio} PIC numerical simulations. The index is here defined by ${\rm d}N/{\rm d}p \propto p^{-s}$. \\
    
    \item While the notion of reference frame is of modest significance in the subrelativistic regime, it becomes crucial in the relativistic regime. The notion of an acceleration timescale, in particular, depends strongly on the frame in which it is calculated. The downstream rest frame, which is about equivalent to the shock rest frame, provides a convenient frame for this purpose.
    
\end{enumerate}

\section{The microphysical view} \label{sec:micr}

When observed on kinetic scales, \emph{i.e.}, of the order of the skin depth $c/\omega_{\rm p}$, where $\omega_{\rm p}=\sqrt{4\pi n_{\rm u} e^2/m}$ represents the plasma frequency of the ambient (far-upstream) plasma ($n_{\rm u}$ being its proper density), the shock front appears as a smooth transition. For reference, $c/\omega_{\rm p} \simeq 1.2\times 10^7\,{\rm cm}\,n_0^{-1/2}$, with the notation $n_x=n/10^x\,{\rm cm}^{-3}$ for the density in $10^x$ cgs units. The description of the inner structure of collisionless shock waves is a long-standing problem of fundamental plasma physics. It was suggested early on that, in the absence of binary collisions, collective electromagnetic modes could account for the dissipation that underpins the shock transition~\cite{Moiseev_1963}. More precisely, such modes are believed to be excited in the shock vicinity through plasma microinstabilities, then to generate a magnetic barrier that slows down, isotropizes and heats up the incoming (unshocked) background plasma.

The relationship between these microphysical processes and the physics of particle acceleration has become clear in recent decades. Gamma-ray burst afterglows, in particular, have provided an important observational test-bed. The modelling of their spectral energy distribution as the synchrotron self-Compton emission of suprathermal distributions of electrons accelerated at the forward relativistic shock, has revealed an effective magnetization\footnote{We distinguish the effective magnetization $\epsilon_B$ in the shock vicinity, from the external magnetization $\sigma$, which pertains to the unshocked ambient plasma. In weakly magnetized shocks, $\epsilon_B$ is generically much larger than $\sigma$ because electromagnetic instabilities, acting in the shock precursor, strongly amplify any pre-existing magnetic energy density.} $\epsilon_B \sim 10^{-3}$ (to within a few orders of magnitude), see {\it e.g.},~\cite{Piran_2004, Lemoine_2013b, Santana_2014}. As the shock is thought to propagate in a medium of low magnetization, $\sigma\sim 10^{-9}$, this effective turbulence must have been self-generated~\cite{Gruzinov_1999, Medvedev_1999}. Additional considerations suggest that such a high effective magnetization must also exist upstream of the shock~\cite{Li_2006}.

On a different level, analytical arguments indicated that the turbulence responsible for relativistic Fermi acceleration must be of a microphysical nature~\cite{Lemoine_2006}, and numerical simulations have confirmed that particle acceleration takes place in the presence of microturbulence generated in the shock precursor~\cite{Spitkovsky_2008a, Nishikawa_2009,Keshet_2009, Martins_2009, Haugbolle_2011, Sironi_2013, Lemoine_2019_PRL}. Furthermore, the instabilities that shape the profile of the shock wave are also those that seed the upstream plasma with an effective magnetization, and which sustain particle acceleration. The accelerated particles, the magnetized turbulence and the shock microphysics thus form an inseparable trio, which explains why one cannot apprehend the acceleration mechanism without discussing in some detail the physics of the instabilities driving the turbulence.

The general picture is then the following. Particles that are energized at the shock front circulate around this shock, and during their acceleration, they execute half-orbits through the background plasma before it crosses the shock front. The extent of the upstream region occupied by these suprathermal particles, called the shock precursor, is determined by their penetration length, which itself depends on their energy. In the rest frame of the background plasma, as it is being swept by the shock precursor, these suprathermal particles form a strongly focused, highly energetic beam, carrying a typical momentum $\sim \gamma_{\rm sh}^2 m$ along the shock normal, with a transverse momentum dispersion smaller by a factor $\sim \gamma_{\rm sh}$. This configuration of two interpenetrating plasmas in the shock precursor begets fast-growing electromagnetic instabilities.

The identification of the dominant modes in the shock precursor is therefore critical to characterizing the nature of the microturbulence, and thus the performance of the acceleration process. We first discuss the case of very weakly magnetized shock waves, generally speaking $\sigma\ll 10^{-4}$, before turning to the case of a more moderate magnetization.

The description that we provide is corroborated by PIC simulations. The PIC method self-consistently evolves the particle distribution coupled to Maxwell's equations~\cite{Birdsall_1991}. Here, the distribution consists of a collection of macroparticles moving through grid-discretized electromagnetic fields. The increasing availability of parallel supercomputers has made this technique widely used to simulate from first principles the kinetics of a broad range of space or laboratory plasmas, whether relativistic or not. However, the coupling between the Lagrangian particles and Eulerian electromagnetic fields at the heart of the PIC method inherently introduces spurious grid-beam instabilities, which are strengthened in the case of a coherent relativistic motion of the particles~\cite{Godfrey_1974, Godfrey_1975}. This configuration is exactly that of the upstream region in relativistic shock simulations, which may then develop nonphysical electromagnetic and plasma modulations, causing artificial particle heating~\cite{Pukhov_1999, Vay_2011}. The simulation capability of PIC codes can be improved by incorporating advanced Maxwell solvers~\cite{Pukhov_1999, Vay_2011, Li_2017} and filtering schemes to attenuate the artificial modes~\cite{Greenwood_2004, Vay_2011, Godfrey_2012, Godfrey_2014}. Obviously, one should be cautious that the latter schemes, while ensuring the apparent stability of the simulated system, do not impair the physical processes considered. Another issue is that PIC simulations are nowadays too computationally demanding to capture the large-scale physics of shock waves in 3D, so that most of them are being conducted in 2D3V (2D in configuration space and 3D in momentum space). Although 2D3V and 3D simulations appear to give overall similar results (at least over scales accessible in 3D)~\cite{Sironi_2013}, differences in the self-generated electromagnetic turbulence have been pointed out~\cite{Haugbolle_2011}, and the impact of a reduced geometry on the long-timescale shock physics is unclear. Moreover, the integration time of state-of-the-art 2D3V simulations is still orders of magnitude below the astrophysical scales, and notably too short for the suprathermal particles to reach a steady-state upstream distribution~\cite{Lemoine_2019_PRL, Lemoine_2019_III}.

A final major difficulty concerns the description of plasmas composed of light and heavy particle species. As of now, the vast majority of simulations have considered shock waves propagating into electron-positron pair plasmas. This choice is mostly dictated by the need to save computational time, as one does not need to resolve the dynamics of the particles with lower mass, as in electron-ion systems. A consequence is that the few existing simulation studies on relativistic electron-ion shocks have all considered strongly reduced ion masses~\cite{Spitkovsky_2008,Martins_2009,  Haugbolle_2011, Sironi_2013, Ardaneh_2015}.
In the case of weak external magnetization, they have revealed that electrons are heated up to near equipartition with the ions, and thus behave as `anti-protons' behind the shock. In the precursor, the electrons follow the dynamics of the ions that carry the bulk of the inertia. Therefore, in some sense, one expects the dynamics of an electron-ion shock to resemble that of an antiproton-proton shock. Further below, we address in more detail how the mass hierarchy of an electron-proton plasma is expected to alter the picture presented in the following.

\subsection{Weibel-mediated shock waves at low magnetization  $\left(\sigma \ll 10^{-4}\right)$} \label{sec:CFImed}

At sufficiently low magnetization levels, the background magnetic field does not affect the instability growth. In this section, we therefore consider an unmagnetized ambient plasma ($\sigma = 0$). The dominant instabilities that arise when a hot dilute beam of particle penetrates a cold denser plasma are the current filamentation instability (CFI),
the electrostatic two-stream and the so-called oblique modes \cite{Bret_2008, Bret_2010a, Bret_2010b, Lemoine_2010, Lemoine_2011,  Rabinak_2011,Shaisultanov_2012}. In the relativistic regime of unmagnetized collisioness shocks, the CFI dominates over most of the precursor, and this is particularly true near the shock front, where the beam-to-plasma density ratio, $n_{\rm b}/n_{\rm p}$, approaches unity. 

Current filamentation instabilities correspond to kinetic-scale instabilities feeding on phase space anisotropies and resulting in the formation of current filaments. One may draw a distinction between the original Weibel instability (WI) \cite{Weibel_1959} and the current filamentation instability~\cite{Fried_1959}. The former arises from thermal anisotropies, while the latter emerges in counterpropagative flow configurations that produce a momentum anisotropy, and which can be regarded as an effective temperature anisotropy. Both have in common the generation of filamentary structures characterized by wavevectors oriented,  respectively, along the cold direction (for the WI) or the normal to the flow (for the CFI). Despite this nuance, the WI and CFI are often referred to interchangeably in the literature, and we shall adopt this convention in later sections. In the context of relativistic shocks, the CFI is triggered by the suprathermal particles counterstreaming against the background plasma in the shock precursor. In the downstream region, where both populations are assumed to be isotropized, these instabilities no longer grow, yet there remains a decaying turbulence \cite{Chang_2008, Keshet_2009, Lemoine_2015} through which the suprathermal particles can radiate \cite{Gruzinov_1999, Rossi_2003, Derishev_2007,Lemoine_2013a,Lemoine_2015b}.

\begin{figure}
    \centering
	\includegraphics[width=0.4\textwidth]{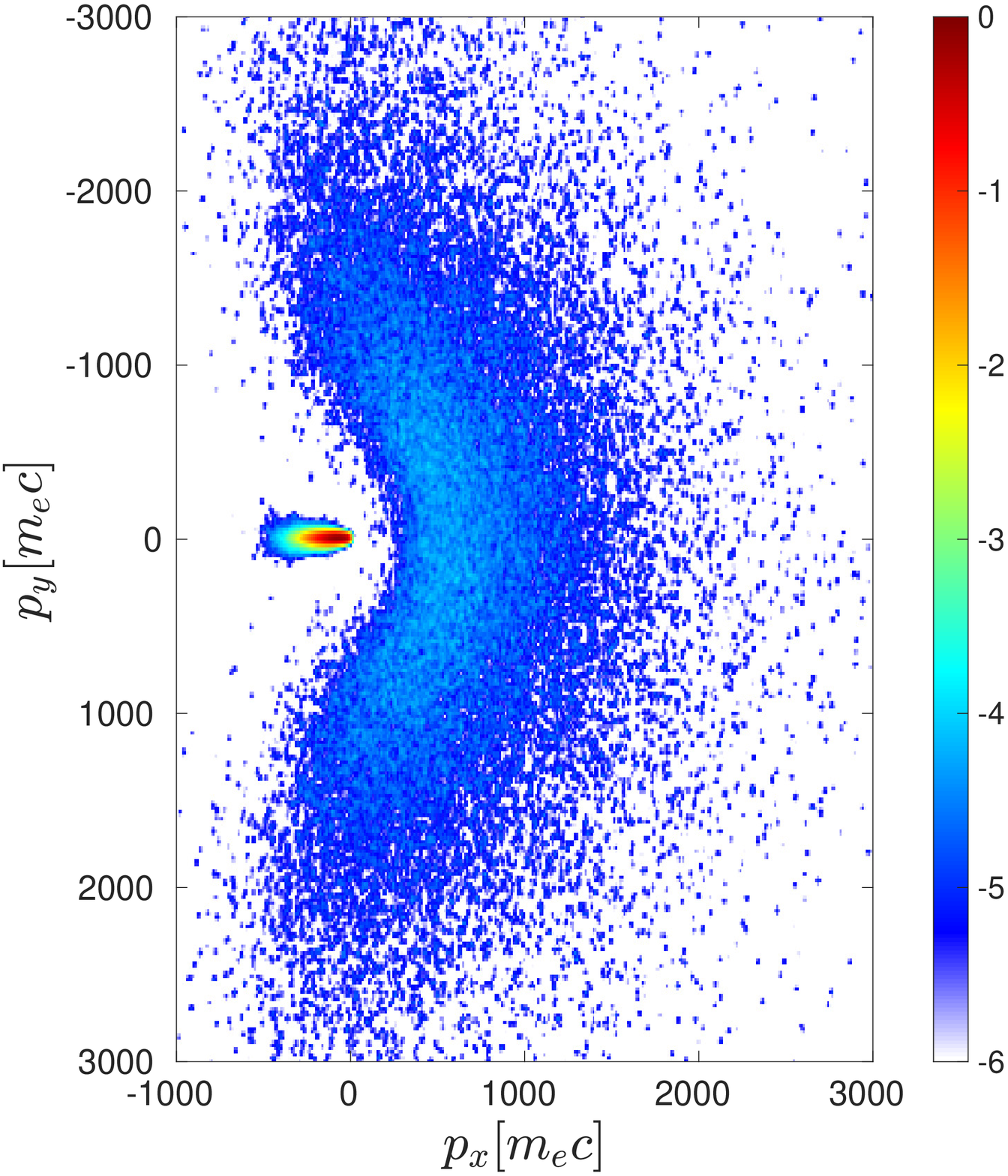}
	\includegraphics[width=0.47\textwidth]{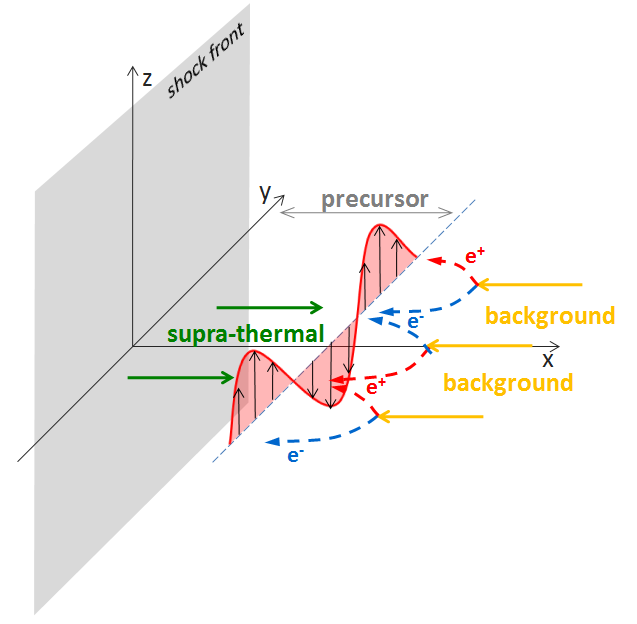}
	\caption{Left panel: Density distribution (in $\log_{10}$ scale, as indicated by the color code) in momentum phase space, in the precursor of an unmagnetized relativistic shock with $\gamma_{\rm sh}=173$ (\emph{i.e.}, $\gamma_{\rm u \vert d} = 100$), as measured in the downstream rest frame of a PIC simulation, close to the shock front. The background plasma corresponds to the compact red (dense and cold) population, moving towards $\boldsymbol{-x}$. The suprathermal particle population, by contrast, appears as a relativistically hot and dilute cloud. Right panel: Mechanism of the CFI induced by the mixing of the background and suprathermal plasmas. The pink harmonic perturbation represents a magnetic perturbation $\delta B_z$ and the dashed lines indicate how the charged species are deflected, thus building up current filaments that amplify the magnetic perturbation. Adapted from~\cite{Pelletier_2017,Vanthieghem_2019}.
	\label{fig:pxpy-CFI} }
\end{figure}

For a concrete illustration, the left panel of Fig.~\ref{fig:pxpy-CFI} shows the particle momentum distribution density (in $\log_{10}$ scale) in a portion of plasma ahead of a relativistic shock front, {\it i.e.}, in the shock precursor. The numerical data is extracted from a 2D3V PIC simulation of an unmagnetized pair shock wave with Lorentz factor $\gamma_{\rm sh} = 173$ \cite{Pelletier_2017,Vanthieghem_2019}. The shock front propagates in the $+x$ direction, see the sketch in the right panel. The phase space -- color code indicating density -- is strongly anisotropic and comprises two distinct populations: the cold (\emph{i.e.}, of small extent in $p_y$) and dense background plasma, drifting with $\beta_x <0$ in the downstream frame $\mathcal{R}_{\rm d}$ (coinciding with the simulation frame), \emph{vs.} the hot and dilute suprathermal particle distribution.

The right panel of Fig.~\ref{fig:pxpy-CFI} sketches the development of the CFI in the shock precursor. Given an initial fluctuation in the transverse magnetic field, $\delta B_z$, here represented as a harmonic mode, the positively and negatively charged components of each (background and suprathermal) plasma population are deflected in different directions, the electrons of one population being pinched together with the positrons of the other around the magnetic-field nodes, and \emph{vice versa}. This results in the formation of alternating-sign current density filaments. This modulated current density exerts a positive feedback on the magnetic perturbation, which further focuses the particles, thus enhancing the current density, etc. The system thus exchanges energy between the reservoir of free energy, associated with the plasma momentum anisotropy, and the magnetic field.

The growth rate of the CFI is large, which makes it the dominant instability in the precursor of weakly magnetized shocks~\cite{Medvedev_1999, Wiersma_2004, Achterberg_2007_I, Achterberg_2007_II, Lemoine_2010, Lemoine_2011, Rabinak_2011, Shaisultanov_2012}. In the rest frame of the background plasma, it reaches $\Im\omega \simeq \omega_{\rm pb}$ in the cold limit, see~\cite{Achterberg_2007_I,Bret_2010a,Lemoine_2010}, where $\omega_{\rm pb}$ denotes the (relativistic) plasma frequency of the suprathermal population,  $\omega_{\rm pb}=\left[4\pi n_{\rm b}e^2/(w_{\rm b}/n_{\rm b})\right]^{1/2}$, with $w_{\rm b}/n_{\rm b}$ the enthalpy per particle. For the suprathermal particle population at a relativistic shock front, $w_{\rm b}/n_{\rm b}\simeq 4 T_{\rm b}$, with $T_{\rm b}=\kappa_{T_{\rm b}}\gamma_{\rm sh}m$  the temperature, and $\kappa_{T_{\rm b}}$ a numerical prefactor for normalization purposes.
The maximum wavenumber for growth is $k_{\rm max} \sim \omega_{\rm p}/c$, meaning that the CFI produces filaments whose typical size is of the order of the background plasma skin depth, that is, at kinetic/microphysical scales.

Although the saturation mechanisms of the CFI are still debated, the magnetic energy density is generally observed to reach a few percents of the available kinetic energy density.

\begin{figure}
    \centering
	\includegraphics[width=0.95\textwidth]{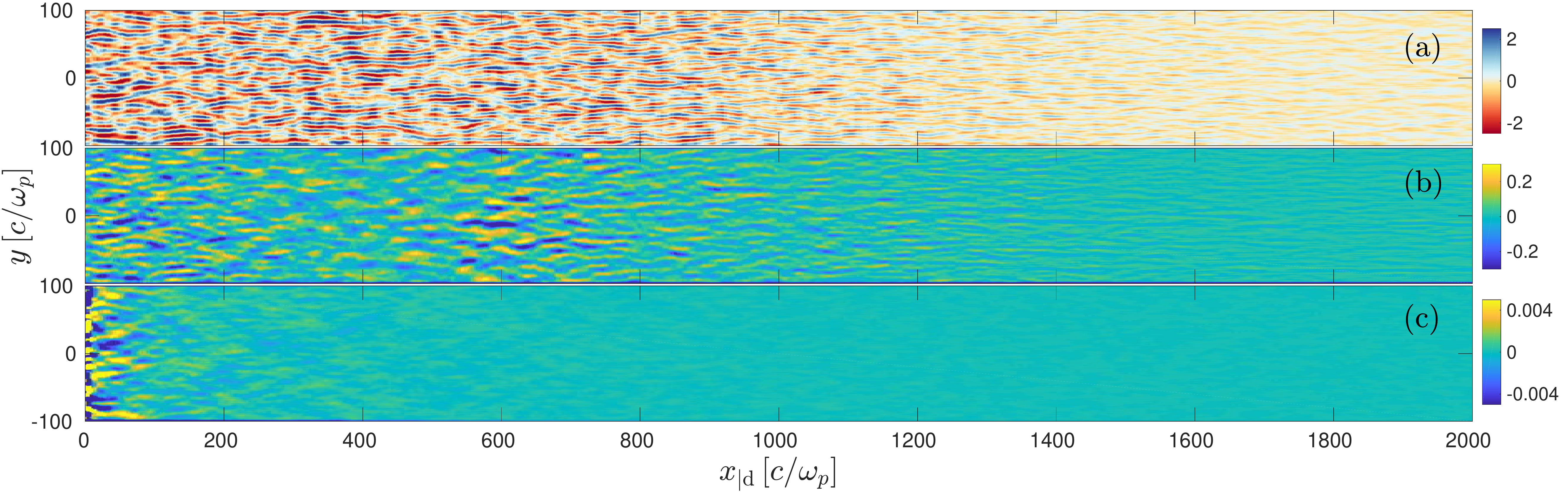}
	\caption{Top: Out-of-plane magnetic field, $B_z$, as observed in a 2D3V pair shock simulation with Lorentz factor $\gamma_{\rm sh}=173$ (\emph{i.e.}, $\gamma_{\rm u \vert d} = 100$) and $\sigma=0$, illustrating the filamentary turbulence generated through the precursor region. The coordinates $x_{\vert\rm d}$ and $y$ are expressed in units of $c/\omega_{\rm p}$, and $x_{\vert\rm d}=0$ indicates the location of the shock front. Middle panel: Density perturbations of the background plasma, $n_{\rm p}(x,y)-\left\langle n_{\rm p}\right\rangle_y$ (average taken over the transverse dimension), as a result of its interaction with the turbulence. Bottom panel: Same for the suprathermal particles, on a scale enlarged by 50 to enhance their weak modulations. From~\cite{Lemoine_2019_III}.
	\label{fig:prec-CFI} }
\end{figure}

As illustrated in Fig.~\ref{fig:prec-CFI}, the shock precursor develops a highly anisotropic spatial structure with filaments of typical width of a few $c/\omega_{\rm p}$ and aspect ratio $\sim \mathcal{O}(10)$. The top panel shows the characteristic geometry of the magnetic field surrounding the current filaments: here in 2D, the out-of-plane ($B_z$) component alternates in sign on each side of a filament. The color code shows how the magnetic field strength increases from the far precursor (to the right) toward the shock. It also reveals the density modulations induced in the background plasma (middle panel), much stronger than those exhibited by the suprathermal particles (bottom panel).

Those profiles can be described analytically, assuming that at each point in the precursor, the system can be approximated as a quasi-steady pressure equilibrium of (transversely) periodic filamentary structures. Consider an infinite, $y$-periodic system of current filaments drifting along the $x$-axis, in (thermal and transverse magnetic) pressure equilibrium \cite{Vanthieghem_2018}. Assuming plasmas of uniform temperatures and drift velocities, then solving for the pressure equilibrium, one finds that the density of species $\alpha$, with charge $q_\alpha$, drift velocity $u_\alpha=\gamma_\alpha\beta_\alpha$, and (proper) temperature $T_\alpha$ can be expressed as
\begin{equation}
    n_\alpha \simeq \langle n_{\alpha}\rangle\left[1 - q_\alpha\frac{\gamma_\alpha  }{T_\alpha}\left(\phi - \beta_\alpha A_x\right)\right]\,.
    \label{eq:nmod}
\end{equation}
This equation assumes weak density modulations, as observed numerically. The electromagnetic potential is here written as $A^\mu = \left(\phi, A_x, 0,0 \right)$. The suprathermal particles have a much larger apparent temperature, $T_{\rm b}/\gamma_{\rm b}$, than the background plasma, hence the above expression predicts a much weaker modulation for this population, in agreement with Fig.~\ref{fig:prec-CFI}. As a matter of fact, the suprathermal particles are so energetic in the frame of the turbulence (to be precised further on), that they can be regarded as rather insensitive to the turbulence, while the background plasma particles are, by contrast, partially trapped in the filaments.

PIC simulations confirm the above theoretical picture of a precursor dominated by a Weibel-type turbulence~\cite{Kato_2007, Spitkovsky_2008, Keshet_2009, Martins_2009, Haugbolle_2011, Sironi_2013, Lemoine_2019_PRL}. In particular, they indicate that the turbulence is mostly magnetostatic in a certain reference frame, which we call the Weibel frame, $\mathcal{R}_{\rm w}$~\cite{Pelletier_2019}. It is expected, because the CFI is itself a magnetic instability, but this, of course, can only be true in one particular frame ($\mathcal R_{\rm w}$). In a configuration where the counterstreaming plasmas share the same characteristics, symmetry considerations dictate that this frame must coincide with the lab frame. 
In the precursor of a relativistic shock, where the interaction is highly asymmetric, the dynamics of this reference frame becomes nontrivial, but proves to be of crucial importance as it sets the frame in which particles undergo elastic interactions in the course of their acceleration.

In the above model of quasi-steady pressure equilibrium, one can characterize the Weibel frame as follows~\cite{Pelletier_2019}. Using Eq.~(\ref{eq:nmod}) in Maxwell's equations and summing over positively and negatively charged species, we derive 
\begin{align}
\partial_y^2 A_{x} & \simeq \frac{4\pi e^2}{m}
\sum\limits_\alpha \langle n_{\alpha}\rangle \gamma_{\alpha} \beta_{\alpha } \left[\frac{\gamma_{\alpha} m}{T_\alpha}
\left(\phi - \beta_{\alpha} A_{x}\right)\right] \,,  &
\partial_y^2 \phi & \simeq \frac{4\pi e^2}{m}\sum\limits_\alpha \langle n_{\alpha} \rangle \gamma_{\alpha} \left[\frac{\gamma_{\alpha}m}{T_\alpha}
\left(\phi - \beta_{\alpha } A_{x}\right) \right] \label{eq:phi_stat_w} \,,
\end{align}
whose solutions reproduce the simulated harmonic pattern of perturbations along the $y$-axis. The Weibel frame is then defined as that in which a solution to the above system can be found with a vanishing electrostatic component ($\phi = 0$), implying 
$n_{\rm b}\gamma_{\rm b\vert w}^2\beta_{\rm b\vert w}/T_{\rm b} = n_{\rm p}\gamma_{\rm p\vert w}^2\beta_{\rm p\vert w}/T_{\rm p}$. Here, quantities indexed by $_{\rm p}$ relate to the background plasma and are understood to be position-dependent inside the precursor. Solving this equation then gives
\begin{align}
    \beta_{\rm w \vert p} &\simeq \frac{\gamma_{\rm p\vert s}^2 }{\kappa^2_{T_{\rm b}}}\xi_{\rm b} \frac{T_{\rm p}}{m} \frac{n_{\rm u}}{n_{\rm p}} \,,&
    \beta_{\rm b\vert w}&\simeq 1\,.
    \label{eq:weibel_frame_speed}
\end{align}
Here $n_{\rm u}$ represents the proper density of the far-upstream (unshocked) background plasma, while $n_{\rm p}$ denotes its proper density at the position considered inside the precursor. The quantity $\xi_{\rm b}$ represents the pressure of the suprathermal particle population in units of $\gamma_{\rm sh}^2 n_{\rm u}m$, the incoming ram pressure of the background plasma in the shock frame. This quantity is a key parameter that characterizes the influence of suprathermal particles in the precursor.
The important lesson to be learned here is that the Weibel frame moves at a subrelativistic velocity relative to the background plasma, and at a relativistic velocity in the shock frame. Hence, the background plasma essentially carries the microturbulence, which embeds the scattering centers of the suprathermal particles.

We draw attention to the fact that the above description holds in pair plasmas. In ion-electron systems, by contrast, the large disparity in inertia between electrons and ions entails differential deceleration between the ions and electrons of the background plasma, together with great differences in their proper density and temperature. Thus, one should distinguish between the background electron and ion contributions in Eq.~\eqref{eq:phi_stat_w}, resulting in an expression for the Weibel frame velocity different from Eq.~\eqref{eq:weibel_frame_speed}.

While the background plasma density (as seen in the shock or downstream frames) remains almost constant all along the precursor (due to its velocity $\vert \beta_x\vert  \sim 1$), the fast decrease in the beam density implies that oblique modes should dominate over the CFI sufficiently far from the shock ($\omega_{\rm p} x/c \gtrsim 10^3$ in the above simulation). These modes are characterized by wavenumbers of oblique orientation with respect to the flow and are of electrostatic nature ($\bm{E}^2-\bm{B}^2>0$). They are expected to develop at the tip of the precursor, with typical wavenumbers $k_x \sim k_y$, and with a maximum growth rate $\Gamma_{\rm max} \simeq \left(\omega_{\rm pb}^{2}\omega_{\rm p}\right)^{1/3}$ \cite{Bret_2010a, Lemoine_2010}.

When counterpropagative plasmas drift at sub-or mildly-relativistic speeds, the system may become dominated by the two-stream instability, associated with wavevectors parallel to the flow, and hence producing longitudinal electrostatic perturbations. Such modes, however, should not play an important role in the relativistic regime.

\subsubsection{Structure of the precursor} \label{sub:shock_structure}
Zooming in on kinetic scales reveals that the shock is composed of a transition region,  which is defined as the zone of strongest gradients in the field and fluid quantities, and which extends over $\sim 100c/\omega_{\rm p}$, preceded by nontrivial dynamics throughout the precursor region, whose size is dictated by the scattering length scale of the high-energy suprathermal particles. 

\begin{figure}
    \centering
	\includegraphics[width=0.8\textwidth]{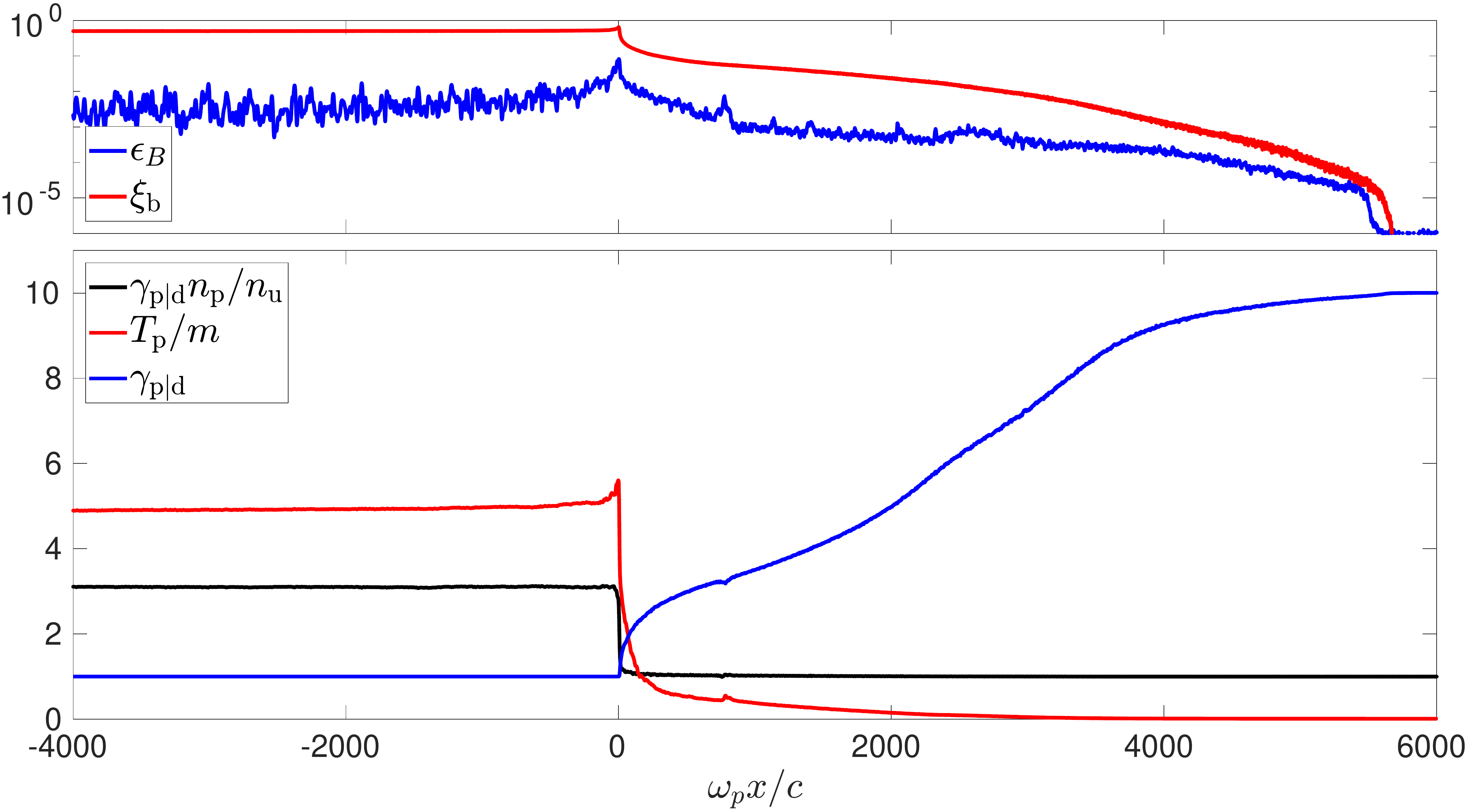}
	\caption{Transversely averaged profiles of the main hydrodynamical quantities in a 2D3V pair shock simulation with initial Lorentz factor $\gamma_{\rm \infty \vert d} = 10$ and temperature $T=10^{-2}\,m_e$. Top panel: beam pressure (red), mean magnetization (blue). Bottom panel: total apparent electronic density in the downstream frame (black), Lorentz factor (blue) and proper temperature (red). Adapted from~\cite{Vanthieghem_2019}.}
	\label{fig:profs}
\end{figure}

This is best illustrated by Fig.~\ref{fig:profs}, which presents a set of one-dimensional (transversely averaged) profiles along the direction of shock propagation, extracted at time $t \simeq 10^4\omega_{\rm p}^{-1}$ from a 2D3V pair shock simulation with Lorentz factor $\gamma_{\rm sh}=17$ (corresponding to relative Lorentz factor $\gamma_{\rm u\vert d} = 10$). The precursor region extends from $x=0$ (the shock front location) to $x\simeq 5\times 10^3c/\omega_{\rm p}$. The background plasma (moving from right to left) is seen to heat up progressively as it decelerates through a microturbulence of growing amplitude, here quantified by the effective magnetization $\epsilon_B = \delta B^2_{\vert \rm s}/\left(4\pi \gamma^2_{\rm sh} n_{\rm u}m c^2\right)$. 

A comprehensive kinetic description of the dynamics of the background plasma in the Weibel-type turbulence has been provided in~\cite{Lemoine_2019_PRL, Pelletier_2019, Lemoine_2019_II}. As its detailed presentation would exceed the frame of this review, we summarize here the salient results, which are important to understand how the PIC simulation results could be extrapolated to the scales of astrophysical interest. 

As a result of their small inertia in the Weibel frame, the background plasma particles are strongly deflected by the magnetic fluctuations, and hence they keep on relaxing in that frame. At the same time, the Weibel frame decelerates (as seen from right to left) in the shock frame because of the growing momentum flux of the suprathermal particles, which interact with the turbulence through scattering. The microturbulence thus serves as a communication agent between the suprathermal and background plasmas. Consequently, since the Weibel frame steadily slows down in the shock frame, in agreement with Eq.~\eqref{eq:weibel_frame_speed}, so does the background plasma. 

Although the background plasma particles experience elastic pitch-angle scattering in $\mathcal R_{\rm w}$, they also gain energy in that frame because of its noninertial nature. This can be seen as some form of Joule heating, wherein the effective gravity plays the role of the driving force, while scattering off the microturbulence acts as collisional friction. This causes the plasma to heat up gradually as it slows down through the turbulence. In this picture, the larger the mean free path of the particles in the turbulence, the faster their heating. Detailed modelling indicates scattering frequencies of $\nu_{\rm | w} \sim 0.01 \omega_{\rm p}$ in the Weibel frame~\cite{Lemoine_2019_II}.

Eventually, as the background plasma comes within a scattering length scale from the shock front, it decouples from the microturbulence, heats up rapidly, to eventually couple back to the microturbulence downstream of the shock, where the conditions are such that the Weibel frame now coincides with the downstream frame. The shock transition has then taken place. In the process, a fraction of the plasma particles are injected into the suprathermal energy tail; thus replenished, the latter population continues fuelling the microinstabilities, thereby sustaining the shock dynamics.

The above physics can be approached through a fluid picture involving the background and suprathermal plasma populations. Because the microturbulence moves subrelativistically with respect to the background plasma, it can be included in the energy budget of the background plasma, and because its contribution to this budget is modest ($\sim 1\,$\%), it can be neglected to leading order. In this description, the shock behaves as a ``cosmic-ray-mediated shock''~\cite{Drury_1981}, since the dynamics of the background plasma in the precursor is controlled by the momentum exchange with the suprathermal population. Such dynamics has been reproduced in nonlinear Monte Carlo simulations, both in the sub- and ultrarelativistic regimes ~\cite{Ellison_2002, Ellison_2004, Ellison_2013, Warren_2015, Ellison_2016}.

Describing the background and suprathermal plasmas as perfect fluids, and solving the equations of conservation of current and energy-momentum in a one-dimensional (along the direction of shock propagation), steady-state configuration, one can derive the following laws of deceleration and heating of the background plasma:
\begin{align}
    &\gamma_{\rm p\vert s} \simeq 1.5\,\xi_{\rm b}^{-1/2}\,,& 
    \frac{n_{\rm p}}{n_{\rm u}}& \simeq n_{\rm u}\,\frac{\gamma_{\rm sh}}{\gamma_{\rm p\vert s}}\,,\nonumber\\
   &\frac{p_{\rm p}}{\gamma_{\rm sh}^2n_{\rm u} m} \simeq 0.15\, \xi_{\rm b}\,,&
    \frac{T_{\rm p}}{m}& \simeq 0.1\,\gamma_{\rm sh}\xi_{\rm b}^{1/2}\,,
    \label{eq:declaw}
\end{align}
where the numerical prefactors have been extracted from the PIC simulations; they agree, to within a factor of the order of unity, with the values predicted by the fluid model. These relationships apply as soon as the dimensionless pressure of the suprathermal particle population, $\xi_{\rm b}$, which has been defined immediately after Eq.~(\ref{eq:weibel_frame_speed}), fulfills $\xi_{\rm b} \gtrsim 1/\gamma_{\rm sh}^2$. At larger distances from the shock (smaller $\xi_{\rm b}$), the background plasma remains in its initial state.

The microturbulence grows slowly in amplitude through the precursor, and it probably reaches a saturated state on long timescales, although this remains the subject of debate. Particle-in-cell simulations indicate that $\epsilon_B$ abruptly rises in the vicinity of the shock front, which has often been attributed to an explosive growth of microinstabilities. On the contrary, the above model suggests that the microturbulence simply grows through adiabatic compression of the magnetic field lines, as $\mathcal R_{\rm w}$ slows down to subrelativistic velocities in the shock transition. From Eq.~\eqref{eq:declaw}, the shock transition occurs when $\xi_{\rm b}$ reaches values $\sim 0.1$, because the plasma has by then been decelerated to mildly relativistic velocities. The value $\xi_{\rm b} \sim 0.1$ can thus be viewed as the typical injection value of the suprathermal particle population in the precursor, and it appears as a key condition for the shock structure. It matches well the fraction of energy density ($\sim 10\,\%$) that has been nearly universally observed in PIC simulations of weakly magnetized, relativistic shocks~\cite{Spitkovsky_2008a, Keshet_2009, Martins_2009, Haugbolle_2011, Sironi_2013, Lemoine_2019_PRL}.
 
\begin{figure}[H]
    \centering
	\includegraphics[width=0.45\textwidth]{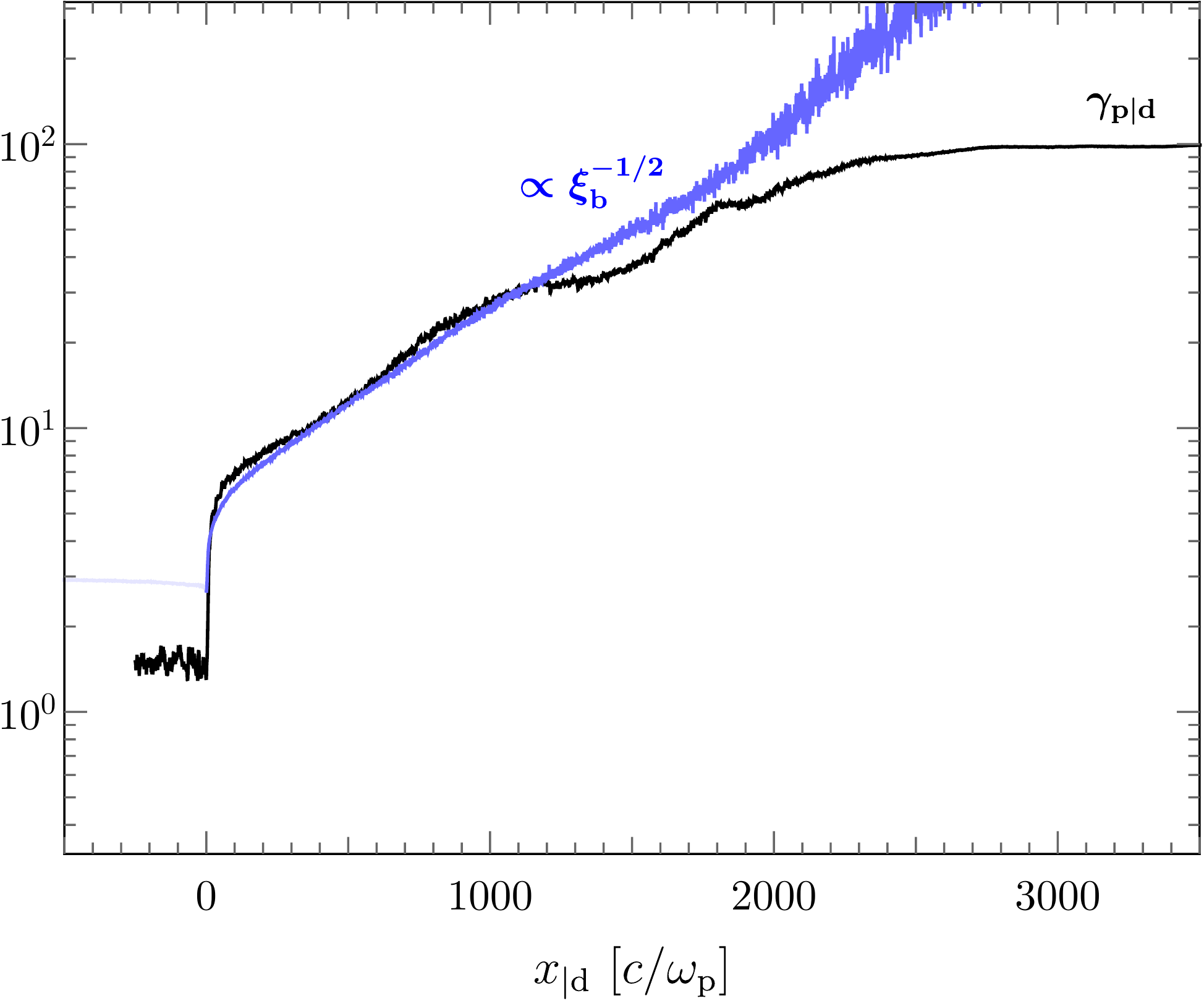}
	\includegraphics[width=0.45\textwidth]{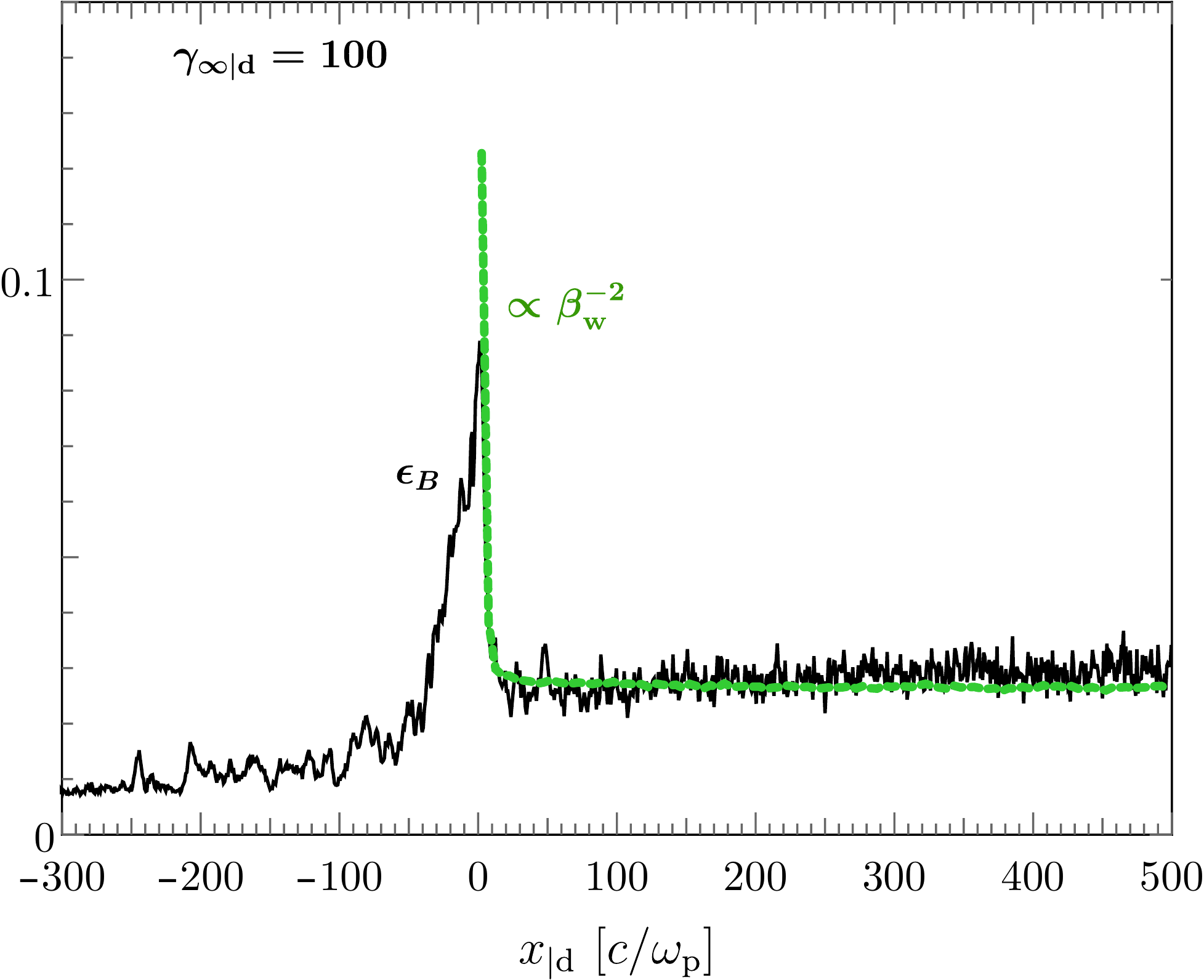}
	\caption{Left panel: Lorentz factor of the background plasma (black curve) from a PIC simulation with $\gamma_{\rm sh}=173$ (\emph{i.e.}, $\gamma_{\rm u\vert d}=100$) and $\sigma=0$. The blue curve plots the theoretical scaling, Eq.~\eqref{eq:declaw}. 
	Right panel: Profile of $\epsilon_B$ around the shock front (black curve), compared with the magnetic compression law, $\epsilon_B \propto \beta_{\rm w\vert s}^{-2}$ (green curve). From~\cite{Lemoine_2019_II}.
	\label{fig:xib_vs_ga}
	}
\end{figure}

To support the above scalings, we plot in the left panel of Fig.~\ref{fig:xib_vs_ga} the evolution of the Lorentz factor of the background plasma across the shock precursor, and juxtapose the scaling law~\eqref{eq:declaw} from the fluid model. Good agreement is observed between the two curves over the plasma deceleration region. The right panel of Fig.~\ref{fig:xib_vs_ga}, which zooms in on the $\epsilon_B$ profile around the shock front, shows that its sudden growth obeys the expected law, $\epsilon_B \propto \beta_{\rm w\vert s}^{-2}$. This close match nicely explains the peak of the magnetic energy at
the shock front, which builds an effective barrier that halts and isotropizes the flow, thereby completing the shock transition.

\subsection{Relativistic shocks of moderate magnetization $\left(10^{-4} < \sigma \ll 10^{-2} \right)$} \label{sec:PCImed}

We now discuss the modification of the shock structure when a mean magnetic field pervades the upstream plasma. We will be interested in magnetizations $\sigma\neq 0$ but $0 < \sigma\ll 10^{-2}$. At the latter upper bound value $\sigma\simeq 0.01$, counterstreaming microinstabilities no longer have time to grow on the crossing timescale of the precursor~\cite{Pelletier_2008,Lemoine_2010, Lemoine_2014a}, and the shock becomes mediated by the compression of the background magnetic field, accompanied by the emission of high-amplitude electromagnetic waves~\cite{Alsop_1988,Gallant_1992,Hoshino_1992,Plotnikov_2019}. The latter case will not be discussed in this article as no strong evidence for Fermi acceleration has been observed in such shocks~\cite{Sironi_2009b,Sironi_2011a}, although see ~\cite{Hoshino_2008, Iwamoto_2019}.

With increasing magnetization, the background (external) magnetic field starts to affect the transport of suprathermal particles, to eventually control the length scale of the precursor at a magnetization $\sigma \gtrsim \sigma_{\rm c}= 10^{-4}$. This critical value can be understood as follows. In the shock rest frame, microturbulent scattering leads to an effective mean free path (scattering length scale)\footnote{More precisely, if $\gamma_{\rm p}$ represents the position-dependent Lorentz factor of the background plasma with respect to the shock front, with $\gamma_{\rm p}=\gamma_{\rm sh}$ outside the precursor, and $\gamma_{\rm p}<\gamma_{\rm sh}$ inside the precursor, as a result of deceleration, then the correct expression for the scattering length is~\cite{Lemoine_2019_III}: $\ell_{\rm scatt\vert s} \simeq \gamma_{\rm p} r_{{\rm g},\delta B}^2/\lambda_{\delta B\vert\rm w}$, with $\lambda_{\delta B\vert\rm w}\simeq c/\omega_{\rm p}$ the turbulence coherence length. The expression in the main text assumes $\gamma_{\rm p}\sim\,$a few, which holds in the \emph{near} precursor, where deceleration to mildly relativistic values has taken place.} $\ell_{\rm scatt\vert s} \sim r_{{\rm g},\delta B}^2/\lambda_{\delta B} \sim \epsilon_B^{-1}(\gamma/\gamma_{\rm sh})^2c/\omega_{\rm p}$
in the near precursor (with $r_{{\rm g},\delta B}=\gamma m/\left(e\delta B\right)$ in terms of the particle Lorentz factor $\gamma$ and the far-upstream background plasma frequency $\omega_{\rm p}$), while the gyration scale in the background magnetic field is simply $r_{{\rm g},B_{\rm ext}\vert\rm s} \simeq \sigma^{-1/2}(\gamma/\gamma_{\rm sh})\,c/\omega_{\rm p}$, with $r_{{\rm g}, B_{\rm ext}\rm\vert s}=\gamma m /\left(e B_{\rm ext\vert s} \right)$ and $B_{\rm ext\vert s}=\gamma_{\rm sh} B_{\rm ext\vert u}$.  This means that at magnetization levels $\sigma \gtrsim \epsilon_B^2\sim 10^{-4}$ (for a representative value $\epsilon_B\sim 10^{-2}$, observed in PIC simulations), the external field effectively governs the transport of all suprathermal particles, because their Lorentz factor verifies $\gamma \gtrsim \gamma_{\rm sh}$, hence $r_{{\rm g},B_{\rm ext}\rm\vert s}\,<\,\ell_{\rm scatt\vert s}$.

This critical magnetization presumably delimits the regime where the physics of the shock is governed by the current filamentation instability ($\sigma \lesssim 10^{-4}$), from that where the external magnetization plays a dominant role ($\sigma \gtrsim 10^{-4}$). In the latter case, the dominant instability is actually supported by this external magnetization, and termed ``perpendicular current driven instability''. We now discuss the structure of the precursor in this magnetization region.

\subsubsection{Perpendicular current driven instability}

In the case where the upstream plasma is magnetized, the precursor size decreases and the rapid advection of the background plasma through the precursor prevents the growth of slow modes~\cite{Pelletier_2008, Pelletier_2009, Lemoine_2010}. In this context, the CFI enters competition with other instabilities which may grow faster during the crossing time of the precursor. Some of these instabilities where discussed in the recent literature, {\it e.g.},~\citep{Bret_2009}. Here, we discuss in some detail the perpendicular current driven instability (PCI), which appears to provide the leading source of magnetization~\cite{Lemoine_2014a, Lemoine_2014b}.

\begin{figure}
\begin{center}
\includegraphics[width=0.45\columnwidth]{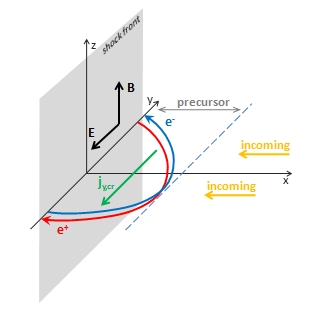}
\includegraphics[width=0.5\columnwidth]{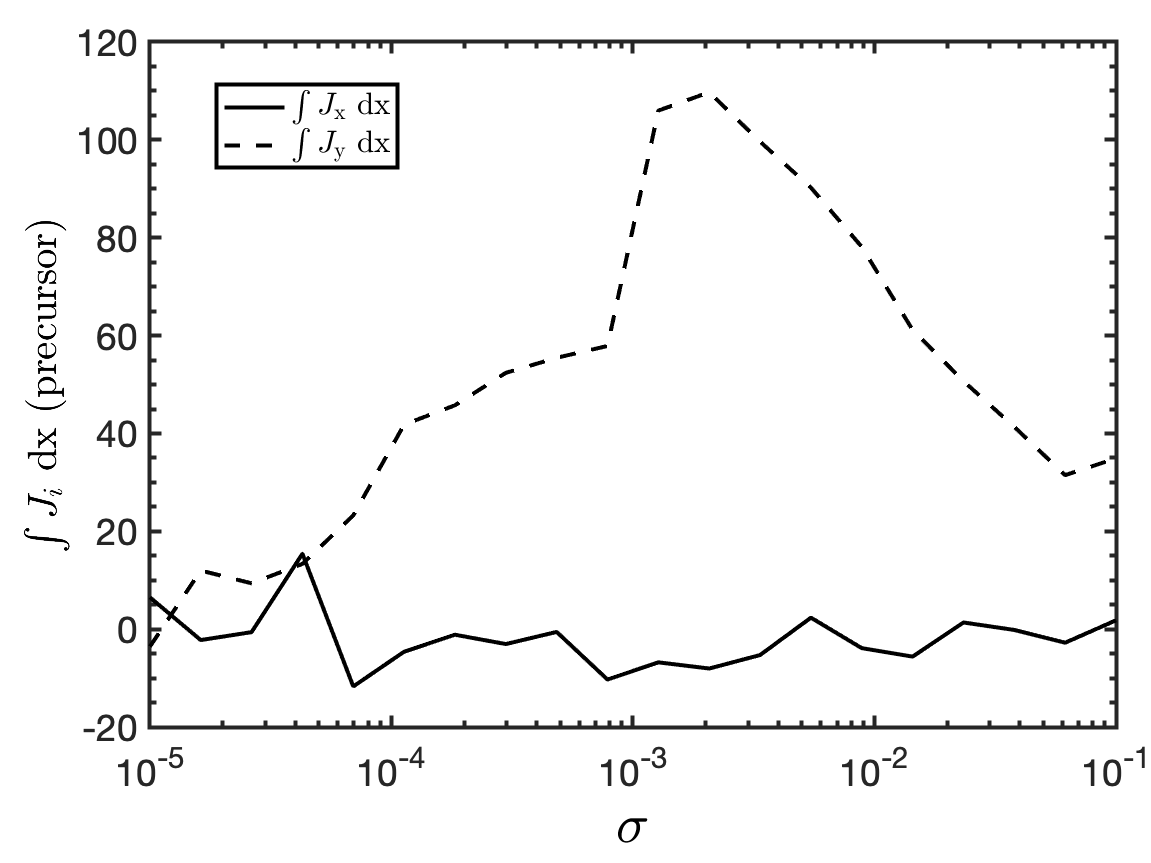}
 \caption{Left panel: Sketch of the development of the perpendicular current driven instability at a moderately magnetized relativistic collisionless shock. Here, the shock propagates toward along the $x$-direction and the external magnetic field lies in $z$-direction. In the shock rest frame, this magnetic field is accompanied by a motional electric field along $-y$. The suprathermal particles gyrate in this electromagnetic structure in the course of their upstream Fermi cycles, generating a net perpendicular electrical current $j_{\rm b}$, oriented along $\boldsymbol{E}$. As the background plasma enters the precursor, it seeks to compensate this perpendicular current, which triggers the PCI.
 Right panel: Measurement of the global $x$ and $y$ electrical, as integrated over the length scale of the precursor, as a function of the external magnetization $\sigma$, taken from PIC simulations reported in \cite{Plotnikov_2018}. At $10^{-4} \lesssim \sigma \lesssim 10^{-2}$, a net perpendicular current is generated. 
  \label{fig:PCI} }
  \end{center}
\end{figure}

In Fig.~\ref{fig:PCI}, we present a sketch of the precursor in this context, illustrating the development of the PCI in the shock front frame (left panel). The typical size of the precursor is $c/\omega_{\rm c}$, with $\omega_{\rm c}=eB_{\rm ext\vert \rm u}/\left(mc\right)$ the upstream cyclotron frequency, $B_{\rm ext\vert \rm u}=B_{\rm ext\vert\rm s}/\gamma_{\rm sh}$ the magnetic field as measured in the unshocked plasma rest frame. As discussed earlier, this length scale also corresponds to the typical gyration radius $r_{\rm g}$ of the suprathermal particles in the shock front rest frame, whose typical Lorentz factor $\sim\gamma_{\rm sh}$.

During their half-Fermi cycles, the suprathermal particles gyrate in the electromagnetic structure, which is composed of $\boldsymbol{B_{\rm ext\vert\rm s}}$ and its accompanying motional electric field $\boldsymbol{E_{\rm \vert s}}=-\boldsymbol{\beta_{\rm p}}\times\boldsymbol{B_{\rm ext\rm \vert s}}$. Particles of opposite charge gyrate in opposite directions along the transverse direction, thereby generating a net electrical current density $\boldsymbol{j_{\rm b}}$, oriented along $\boldsymbol{E_{\rm \vert s}}$. The diamagnetic effect of this current is so large, that it has to be compensated in a steady state situation.

The magnitude of this current is $j_{\rm b}\,\sim\,\gamma_{\rm sh}\xi_{\rm b}n_{\rm u}e$. The number density of suprathermal particles in the shock frame is indeed of the order of $\gamma_{\rm sh}\xi_{\rm b}n_{\rm u}$, because -- omitting numerical prefactors for clarity -- they carry a fraction $\simeq\xi_{\rm b}$ of the downstream energy density $\sim \gamma_{\rm sh}^2 n_{\rm u} m$  with a typical energy per suprathermal particle $\sim\gamma_{\rm sh}m$~\cite{Lemoine_2014a}. As the particles of the unshocked plasma enter the precursor, they are driven in the perpendicular direction to achieve current compensation. Positrons drift towards $\boldsymbol{+y}$ while electrons drift towards $\boldsymbol{-y}$. In this shock frame, the background plasma is inflowing with Lorentz factor $\gamma_{\rm sh}$ and apparent density $\gamma_{\rm sh}n_{\rm u}$, hence the transverse 3-velocity is of the order of $\xi_{\rm b}$, which implies a perpendicular four-velocity $\vert u_y\vert \sim \gamma_{\rm sh}\xi_{\rm b}$. Consequently, $\vert u_y\vert\,\gtrsim\,1$ is expected for relativistic shocks, possibly $\vert u_y\vert\,\gg\,1$.

Figure~\ref{fig:PCI} (right panel) shows the PIC-simulated electrical currents flowing through the precursor of relativistic shocks with $\gamma_{\rm sh}=17$, as a function of magnetization. A net perpendicular current is indeed observed at magnetizations $10^{-4}\lesssim \sigma \lesssim 10^{-2}$, in agreement with the above estimates~\cite{Plotnikov_2018}.

As detailed in Refs.~\cite{Lemoine_2014a,Lemoine_2014b}, the PCI can be regarded as a form of filamentation instability acting in the transverse direction, and creating filaments of non-alternating polarity. Neglecting indeed the response of the suprathermal particles to the growth of the instability, this latter can be pictured as follows. As the oppositely charged species of the background plasma flow into opposite directions along $\boldsymbol{y}$, they are focused into the nodes of any perturbation $\delta B_z$ modulated in $x$, as for the filamentation instability (in which particles inflowing along $\boldsymbol{x}$ are pinched into filaments modulated in $y$ by a perturbation $\delta B_z$). One difference, here, is that the currents do not alternate in polarity, because of the existence of a net current carried by the background plasma. The PCI breaks up this uniform current into current filaments of a same polarity. 

A relativistic two-fluid analysis of this instability indeed confirms that the fastest growing mode exhibits a wavenumber oriented along the shock normal ($x$ direction), and dedicated PIC simulations have confirmed the fast growth of this instability, with $\Im\omega\simeq\vert\beta_y\vert\omega_{\rm p}$, with $\vert\beta_y\vert \simeq\,{\rm min}\left(1,\gamma_{\rm sh}\xi_{\rm b}\right)$ the transverse 3-velocity~\cite{Lemoine_2014a}. The growth rate can thus be as fast as $\omega_{\rm p}$, which exceeds that of the CFI.

\subsubsection{Structure of the precursor}

The deflection of the incoming flow along $\boldsymbol{y}$ implies a substantial deceleration of the flow along $\boldsymbol{x}$. One can apprehend this slowdown as follows. The Lorentz factor of the flow remains large, in particular the total 3-velocity $\vert\boldsymbol{\beta}\vert\sim 1$, up to corrections of order $\gamma_{\rm sh}^{-2}$. However, conserving the energy per particle while achieving current compensation implies that $\beta_{\rm x}$ deviates from its initial value by quantities of the order of $\xi_{\rm b}^2$. In other words, the plasma decelerates to longitudinal bulk Lorentz factor $\gamma_{\rm p}\,\sim\,\xi_{\rm b}^{-1}$ if $\xi_{\rm b}\,\gg\,1/\gamma_{\rm sh}$, but retains its initial asymptotic value of $\gamma_{\rm sh}$ if $\xi_{\rm b}\,\ll\,1/\gamma_{\rm sh}$.

Alternatively, one can model this using a fluid model,  as for the CFI-mediated shocks discussed in Sec.~\ref{sec:CFImed}, attributing the transverse motion to an effective temperature, given that positively and negatively charged species flow into opposite directions. This temperature is thus of magnitude $T_{\rm p} \sim \vert u_y \vert m$. Conservation of particle current density imposes $n_{\rm p}\gamma_{\rm p}\beta_{\rm p}=n_{\rm u}\gamma_{\rm sh}\beta_{\rm sh}$ in steady state in the shock frame, while assuming conservation of energy density gives $\gamma_{\rm p}^2\beta_{\rm p}w_{\rm p}=\gamma_{\rm sh}^2\beta_{\rm sh}n_{\rm u}m$. Here, $w_{\rm p}$ (resp. $n_{\rm p}$) denotes the enthalpy (resp. number) density at a given location, as before; the plasma is initially cold. We thus derive $\gamma_{\rm p}=\gamma_{\rm sh}/\left(w_{\rm p}/n_{\rm p}\right)$, hence $\gamma_{\rm p}\sim \xi_{\rm b}^{-1}$ for $\vert u_y\vert\simeq \gamma_{\rm sh}\xi_{\rm b}\gg 1$, once the plasma becomes relativistically hot ($w_{\rm p}/n_{\rm p}\simeq 4T_{\rm p}$).

This is a quite remarkable feature: the compensation of the current slows down the incoming plasma, such that, at large values of the current, $\gamma_{\rm sh}\xi_{\rm b}\,\gg\,1$, the relative Lorentz factor between the plasma and the shock front becomes of the order of $1/\xi_{\rm b}$, independent of the initial  Lorentz factor. In this sense, the shock precursor plays the role of a buffer.  While for $\sigma\lesssim10^{-5}$, the slowdown of the incoming plasma is ensured by the scattering of suprathermal particles off the microturbulence, it results here from current compensation. While for $\sigma \lesssim 10^{-5}$, $\gamma_{\rm p}\propto\xi_{\rm b}^{-1/2}$ and $\xi_{\rm b}\propto x^{(2-s)/2}$ for a power-law distribution ${\rm d}N/{\rm d}p\propto p^{-s}$~\cite{Lemoine_2019_III}, here we have $\gamma_{\rm p}\propto\xi_{\rm b}^{-1}$ and $\xi_{\rm b}\propto x^{2-s}$. Hence, when the transport is dominated by the external magnetic field, the background plasma evolves more rapidly in the precursor.

\subsubsection{A view in terms of magnetization}

\begin{figure*}
	\begin{center}
		\includegraphics[width=0.9\textwidth]{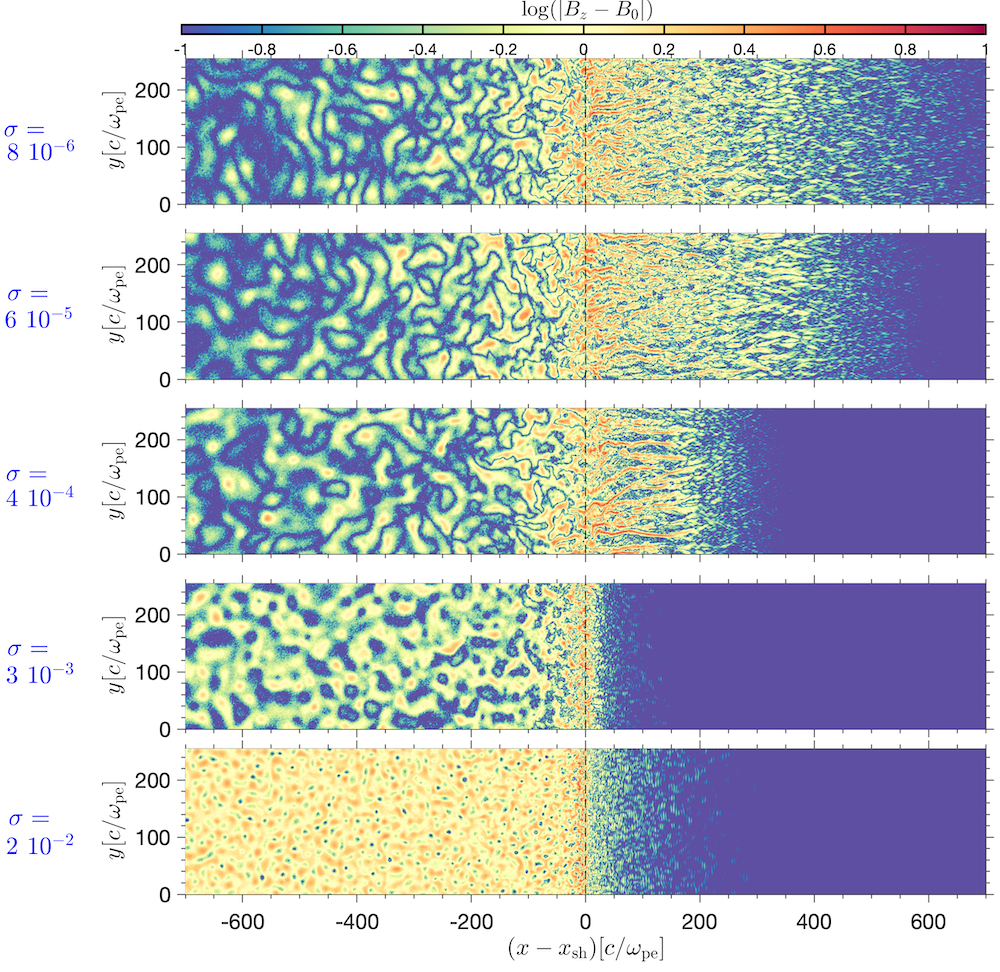}
		\caption{Evolution of the shock layer structure with increasing magnetization of the ambient medium, from 2D3V PIC simulations of a pair shock with Lorentz factor $\gamma_{\rm sh}=17$. Each panel displays the magnetic field enhancement in the simulation plane, $\ln [{e\vert B_z-B_0\vert} /(m c \omega_{\rm pe})]$ (or, equivalently $\ln [\gamma_{\rm u\vert d}\epsilon_B^{1/2}]$). Five representative magnetizations are presented, from top to bottom: $\sigma=8\times 10^{-6}, 6\times 10^{-5}, 4\times 10^{-4},3\times10^{-3}$, and $2\times10^{-2}$. Adapted from \citep{Plotnikov_2018}.} 
		\label{fig:structure_sigma}
	\end{center}
\end{figure*}

Figure~\ref{fig:structure_sigma} presents (in $\ln$ scale) the magnetic field strength $e\vert B_z-B_0\vert /(m c \omega_{\rm pe})$ (or, equivalently, $\gamma_{\rm u\vert d}\epsilon_B^{1/2}$) as obtained from PIC simulations at $\gamma_{\rm sh}=17$ and different magnetizations. This figure is adapted from Ref.~\cite{Plotnikov_2018}, but see also~\cite{Sironi_2013}. At the top, corresponding to low magnetizations ($\sigma< 10^{-4}$), the turbulence takes on a filamentary structure, elongated along the shock normal, typical of the CFI. Accordingly, at those magnetizations, the net perpendicular current shown in Fig.~\ref{fig:PCI} is weak, because the incoherence of the microturbulence dominates the regular pattern of gyration in the background magnetic field. Hence, the shock precursor is dominated by the CFI. In the bottom panels, for $\sigma >10^{-3}$, the size of the precursor is too restricted to perform a detailed study of the microturbulence, but its structure observed downstream suggests that a different mechanism is at play and indeed, the perpendicular current is strong in that case. The PCI presumably dominates the generation of turbulence in this case.

The shortening of the precursor with increasing magnetization is a nontrivial effect that deserves some comments. The maximal extent of the precursor is set by the penetration length of the highest energy particles, but its typical extent is set by the penetration depth of the bulk of suprathermal particles, with Lorentz factors $\gamma \sim \gamma_{\rm sh}$. At intermediate magnetizations $\sigma \gtrsim 10^{-4}$, the background magnetic field controls the trajectories of the particles upstream of the shock, hence the typical length scale is the gyroradius $r_{\rm g}\simeq \sigma^{-1/2}c/\omega_{\rm p}$, and the precursor scale indeed diminishes with increasing $\sigma$. At lower magnetizations, $\sigma \lesssim 10^{-4}$, the microturbulence rather determines the penetration length, which becomes $\ell_{\rm scatt | d} \sim \epsilon_B^{-1}c/\omega_{\rm p}$, independent of $\sigma$. 

The actual appearance of the shock is somewhat modified by the fact that, as particles are accelerated to higher and higher energies, their penetration depth increases, thereby enlarging the precursor. The appearance of the precursor also depends on the scaling of the various parameters ($\xi_{\rm b}$, $\gamma_{\rm p}$, etc.) with $x$, which differs at low and intermediate magnetizations. Furthermore, the maximal energy of accelerated particles itself depends on the magnetization, at intermediate magnetizations at least. This will be addressed in the forthcoming section. 

Finally, for $\sigma > 10^{-2}$ the microturbulence does not have time to grow in the precursor, as mentioned above. It is replaced by large-amplitude electromagnetic waves emitted from the shock front, which result from maser emission of particles gyrating synchronously in the compressed background magnetic field~\cite{Pelletier_2008,Lemoine_2010, Lemoine_2014a}. 

\section{Phenomenological consequences}\label{sec:phen}

In this section, we summarize the various phenomenological implications of the microphysical picture derived above, as well as their potential observables in high-energy astrophysics.

\subsection{Acceleration and spectral index}

We first recall that, at the magnetizations that we are interested in ($\sigma \lesssim 10^{-3}$), particle acceleration can take place at relativistic shock waves, although it is restricted to a finite range of gyroradii (notwithstanding other limitations associated to energy losses and escape), see~\cite{Sironi_2013} for simulations. This can be understood as follows. In the relativistic regime, the magnetic configuration is most generically superluminal, hence the background magnetic field inhibits acceleration by dragging the particles away from the shock front, at the velocity $\beta_{\rm s\vert d}\simeq 1/3$. At those magnetizations, however, the precursor becomes permeated by an intense microturbulence, which can unlock the particles from the background field lines, and thus sustain acceleration~\cite{Lemoine_2010}. This happens provided the scattering frequency of particles in this (downstream) microturbulence, $\nu_{{\rm scatt},\delta B} \sim \epsilon_B\,\left(\gamma/\gamma_{\rm sh}\right)^{-2}\,\omega_{\rm p}$, exceeds their gyrofrequency in the background field, $\nu_{\rm c,B_{\rm ext}}\sim \sigma^{1/2}\left(\gamma/\gamma_{\rm sh}\right)^{-1}$~\cite{Pelletier_2009}. Since $\nu_{{\rm scatt},\delta B}$ falls off faster with energy than $\nu_{\rm c,B_{\rm ext}}$, this implies the existence of a maximal (downstream frame) Lorentz factor, $\gamma_{\rm max} \sim \epsilon_B \sigma^{-1/2}\gamma_{\rm sh}$, above which the particles exhibit such a large gyroradius, as compared to the typical scale of the microturbulence, that their scattering becomes so feeble that they are more rapidly advected away from the shock front by the background field than scattered back to the shock by the microturbulence. This scaling has been tentatively detected in the PIC simulations reported in ~\cite{Plotnikov_2018}. Using their results, the dependence of this maximal Lorentz factor can be scaled as
\begin{equation}
\gamma_{\rm max} \approx 0.5\,\sigma^{-1/2} \gamma_{\rm sh}\,.
\label{eq:gmax}
\end{equation}
The above holds for a shock propagating in a pair plasma. In an electron-proton plasma, $\gamma_{\rm max}$ is multiplied by $m_p/m$, where $m$ represents the mass of the accelerated particle. At $\sigma \sim 10^{-3}$, the powerlaw distribution can already extend over $\sim 1$ order of magnitude beyond the injection energy, leading to a synchrotron spectrum covering 2 decades~\cite{Plotnikov_2018}.

A well-known prediction for the slope of the accelerated population is $s\simeq 2.2$ in the ultra-relativistic regime $u_{\rm sh}\gg1$ (recalling that the slope is defined by ${\rm d}N/{\rm d}p\propto p^{-s}$)~\citep{Bednarz_1998, Kirk_2000, Achterberg_2001, Ellison_2002, Lemoine_2003, Niemiec_2004, Keshet_2005, Warren_2015}. This estimate, which has been obtained through test-particle Monte Carlo simulations and analytical calculations, assumes that the particles can scatter isotropically, meaning with equal probability in all directions, in the turbulence rest frame. Since both the mean magnetic field and the shock normal set privileged directions, this may not always be the case. In the presence of anisotropic scattering, the spectral index is typically steeper, with $s\simeq 2.4-2.7$, {\it e.g.}~\cite{Lemoine_2006a,Niemiec_2006,Keshet_2019}.

\begin{figure}
\begin{center}
\includegraphics[width=0.85\columnwidth]{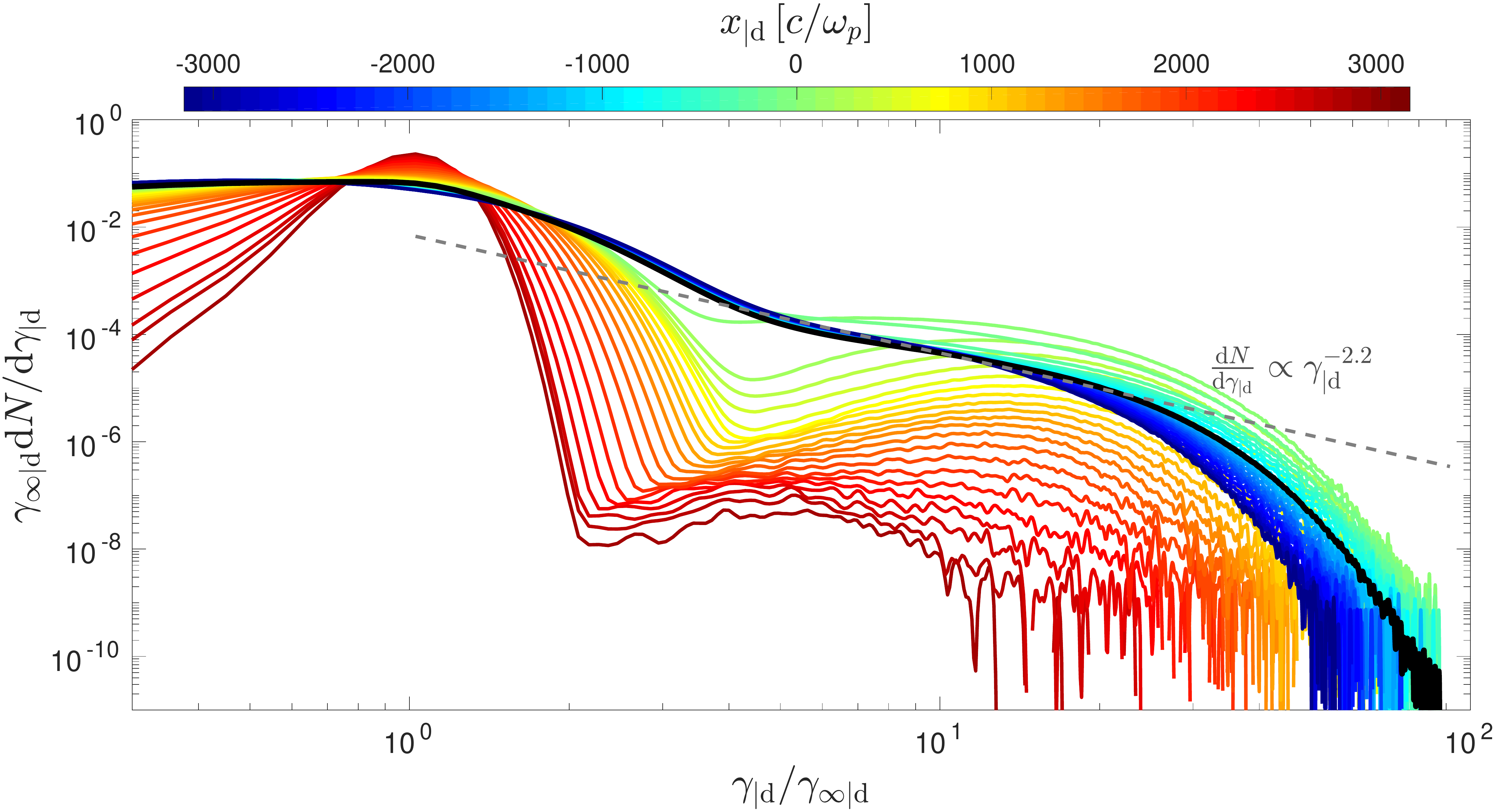}
 \caption{Particle energy spectrum taken from a PIC simulation of a relativistic pair shock with $\gamma_{\rm sh}=173$ and $\sigma=0$, at different distances $x_{\vert\rm d}$ from the shock front, as indicated by the color scale. Positive (resp. negative) values of $x_{\vert\rm d}$ correspond to upstream (resp. downstream). The black line shows the spectrum integrated over the whole simulation box and is compared by the theoretical prediction of a powerlaw with $s=2.2$. From~\cite{Lemoine_2019_III}.
  \label{fig:spec_CFI} }
\end{center}
\end{figure}

A direct measurement of the slope of the accelerated particle population can be carried out in PIC simulations, although the restricted range over which this population is obtained, as a consequence of both the above intrinsic limitation associated with the magnetization and the relatively short time over which such simulations can be conducted, restrains its accuracy. Furthermore, this slope may be affected by the geometry of the simulation, which nowadays is typically 2D3V for the larger-scale computations~\cite{2020arXiv200211123L}. The observed slope is $s \simeq 2.3 \pm 0.1$ at $u_{\rm sh}\gg 1$~\cite{Spitkovsky_2008a,Nishikawa_2009,Keshet_2009, Martins_2009, Haugbolle_2011, Sironi_2013}. Figure~\ref{fig:spec_CFI} shows the energy spectrum of accelerated particles at different distances from the shock in the unmagnetized case ($\sigma=0$). It is worth noting that similar trends are observed for $\sigma \in \left[ 10^{-6}, 10^{-3} \right]$~\cite{Plotnikov_2018}.

\subsection{Acceleration rate and consequences on the particle maximum energy}

The maximal energy can be further limited by energy losses and/or escape losses, the former applying in general to electrons and positrons, the latter to ions. The limiting energy then depends on the acceleration rate. Although phenomenological studies often rely on the Bohm estimate $t_{\rm acc}\sim r_{\rm g}/c$, the latter does not apply to weakly magnetized relativistic shock waves, because the small-scale nature of the turbulence slows down the acceleration: the larger the energy of the particle, the larger the ratio of its gyroradius to the length scale of the turbulence, hence the less effective the scattering, and the more $t_{\rm acc}$ departs from the Bohm scaling. For the correct estimate of the scattering time up to a prefactor $\kappa$ (see thereafter), recalled above, we obtain~\cite{Pelletier_2009,Lemoine_2010,Plotnikov_2013}
\begin{equation}
    t_{\rm acc} \approx \kappa\,\left(\mu\frac{\gamma}{\gamma_{\rm sh}}\right)^2\,\epsilon_B^{-1}\,\lambda_{\delta B}\,,
    \label{eq:tacc}
\end{equation}
where $\mu$ represents the ratio of the mass of the accelerated particle to the mass of the inertia carriers, meaning $\mu=m_e/m_e=1$ for a pair shock, $\mu=m_e/m_p\simeq 1/1836$ for an electron in an electron-proton shock,  $\mu=m_p/m_p=1$ for a proton in an electron-proton shock, etc., and $\lambda_{\delta B}$ the coherence length of the microturbulence. Here, we adopt $\epsilon_B \simeq 10^{-2}$, as observed in PIC simulations. 

\begin{figure}
	\begin{center}
\includegraphics[width=0.9\columnwidth]{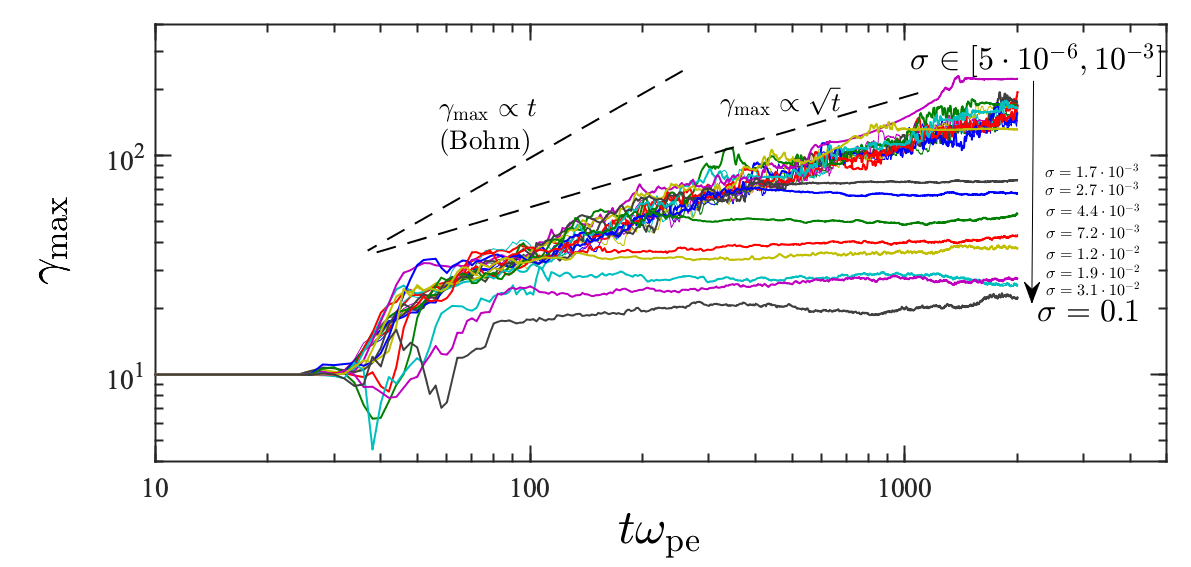}
 \caption{Temporal evolution of the maximal Lorentz factor of the  particle distribution in different simulations with increasing magnetization (lines with different colors). The dashed lines indicate, for reference, the $\gamma_{\rm max} \propto t$ (Bohm regime), and $\gamma_{\rm max} \propto \sqrt{t}$ scaling (small-angle scattering). From Ref.~\citep{Plotnikov_2018}.}
   \label{fig:accrate} 
   	\end{center}
\end{figure}

The numerical prefactor $\kappa$, of the order of unity, can be determined from PIC simulations, using the fact that Eq.~(\ref{eq:tacc}) above implies $\gamma_{\rm max}(t)\simeq\kappa^{-1/2}\gamma_{\rm sh}\mu^{-1}\epsilon_B^{1/2}\left(2 t\omega_{\rm p}\right)^{1/2}$, which can be confronted with the scaling of the maximal Lorentz factor in PIC simulations, reported in Fig.~\ref{fig:accrate} (for a pair shock). From this figure, we derive $\kappa \simeq 0.4$.

Comparing the acceleration timescale to the synchrotron loss timescale, $t_{\rm syn} = 6\pi m_e c/\left(\sigma_{\rm T}\delta B^2\gamma\right)$, then allows one to express the \emph{electron} maximal Lorentz factor in a simple way, as a function of the turbulence length scale and $r_e$ the classical electron radius:
\begin{equation}
    \gamma_{e,\,\rm max} \simeq  \left(\frac{9}{4\kappa}\frac{\lambda_{\delta B}}{r_e}\right)^{1/3} \simeq 2.0\times 10^4 \kappa^{-1/3}\lambda_{\delta B,0}^{1/3}\,,
    \label{eq:gmax2}
\end{equation}
with $\lambda_{\delta B,0}$ the turbulence length scale in cm. For instance, if $\lambda_{\delta B}\simeq 10c/\omega_{\rm p}$, and $c/\omega_{\rm p}\simeq 10^7\,\rm cm$, corresponding to a density (of the unshocked electron-proton plasma) of $1\,\rm cm^{-3}$, the electron maximal Lorentz factor is of the order of $10^7$~\cite{Plotnikov_2013}. This value is calculated in the reference frame of the blast (the shocked plasma).  In the early phase ($\lesssim 100\,\rm s$) of a gamma-ray burst afterglow, this leads to the emission of synchrotron photons with an energy as large as $\sim1\,\rm GeV$, {\it e.g.}~\cite{Plotnikov_2013, Wang_2013, Sironi_2013}, as seemingly required by the observation of extended high-energy emission in gamma-ray bursts~\cite{Naval_2018}.

For protons, the limiting constraint rather derives from the age of the shock, $R/(\gamma_{\rm sh}\beta_{\rm sh}c)$, in which case one obtains~\cite{Plotnikov_2013}
\begin{equation}
    E_{\rm p,\,max} \simeq 1\,{\rm PeV}\,R_{17}\,\left(\gamma_{\rm sh}/100\right)^{3/2}\,n_0^{1/4}\,.
    \label{eq:epmax}
\end{equation}
Here, $E_{\rm p,\,max}$ is calculated in the source rest frame, thus including a boost by a Doppler factor $\delta_{\rm D}\simeq\gamma_{\rm sh}$.

\subsection{Minimum electron Lorentz factor in electron-proton plasmas}

In the rest frame of a weakly magnetized relativistic shock, the energy reservoir is carried by the incoming background plasma, under the form of kinetic energy. Consequently, in an electron-proton shock, the electrons carry a fraction $m_e/m_p$ less energy than the ions, and if both species were to satisfy the hydrodynamical shock crossing conditions separately, then behind the shock, the electrons would still carry a small fraction $\epsilon_e \simeq m_e/m_p$ of the energy density. Particle-in-cell simulations have shown that, on the contrary, behind the shock, $\epsilon_e \simeq 0.3$, meaning that the electrons have drawn energy from the initial ion kinetic energy reservoir~\cite{Spitkovsky_2008,Sironi_2011a,Haugbolle_2011,Sironi_2013}. 

Equivalently, they have been heated up to a large minimum Lorentz factor, $\gamma_{e,\rm min} \sim \epsilon_e\,\gamma_{\rm sh}\,m_p/m_e$, implying a mean energy close to that of the protons, for which $\gamma_{p,\rm min} \sim \gamma_{\rm sh}$. A value $\epsilon_e\simeq0.1-0.3$ appears in rather good agreement with most light curves of gamma-ray burst afterglows, {\it e.g.,}~\cite{Piran_2004,Kumar_2015a,van_Eerten_2018}. 

In the frame of the microphysical model developed in the previous sections, this energy transfer is the consequence of the different inertias of the background electrons and protons: as these species relax at different rates in the Weibel frame $\mathcal R_{\rm w}$, the resulting charge separation induces an electrostatic potential that pulls the electrons through the turbulence, in line with the ions, enhancing electron friction and therefore heating. We point out that other mechanisms have been proposed in the literature, {\it e.g.},~\cite{Medvedev_2006, Kumar_2015}, and that the detailed physics of heating remains debated.

It is also interesting to note that phenomenological models of extreme blazars, defined as those whose observed synchrotron peak energy $\epsilon_{\rm syn,\,max} \gtrsim 1\,{\rm keV}$, imply a large minimum Lorentz factor of the electron distribution, $\gamma_{e,\rm min}\gtrsim 10^3$, as well as a low magnetization $\sigma\,\lesssim\,10^{-3}$, see {\it e.g.}~\cite{Biteau_2020}. It is thus tempting to interpret this observation as the signature of electron preheating in a relativistic electron-proton shock, as described here.

\subsection{Fate of downstream turbulence}

\begin{figure}
	\begin{center}
 \includegraphics[width=0.9\columnwidth]{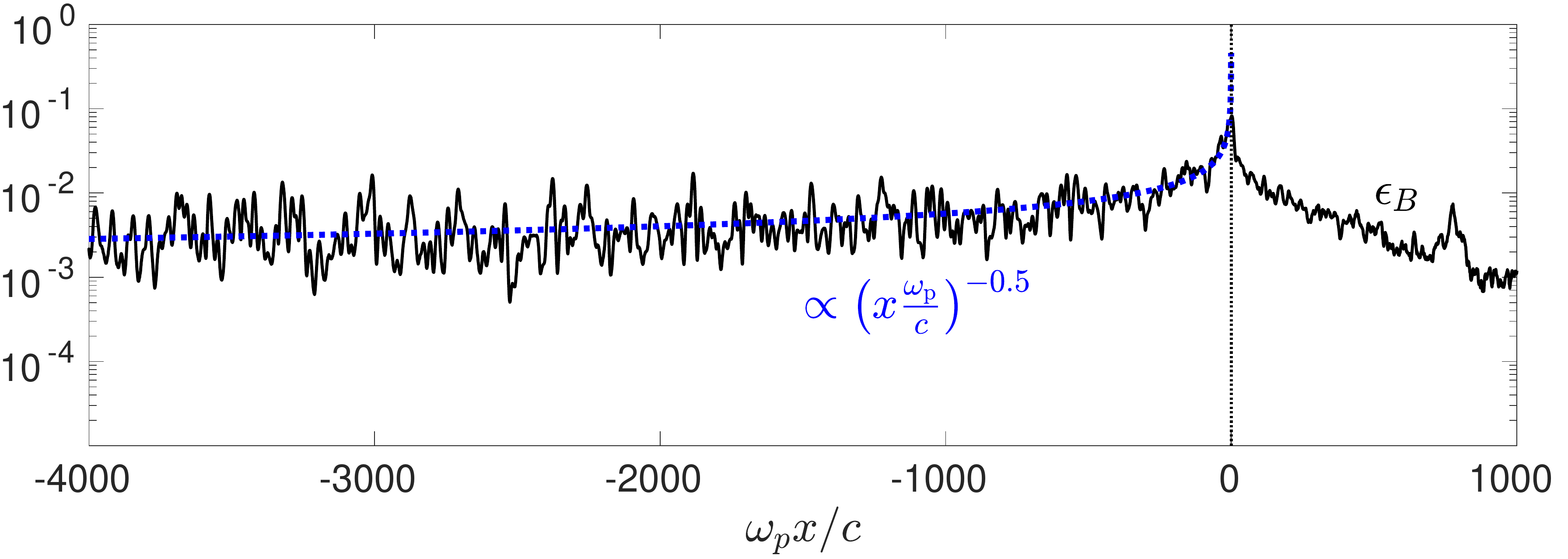}
 \caption{Decay of the microturbulence in the shocked region, as observed in a PIC simulation for $\gamma_{\rm sh}=17$. The dashed blue line shows a law $\epsilon_B \propto \left(x\omega_{\rm p}/c\right)^{-0.5}$.}
   \label{fig:mudecay} 
   	\end{center}
\end{figure}

Another important consequence of the shock microphysics is that the turbulence, which sustains the acceleration process, exhibits a typical length scale of the order of the plasma skin depth $\sim c/\omega_{\rm p}$, hence it is prone to decay through phase mixing. Figure~\ref{fig:mudecay} shows the spatial decay law observed in a PIC simulation of a pair shock with $\gamma_{\rm sh}=17$.

Phase mixing erodes the magnetic fluctuations by erasing the small-scale structures first, with a damping rate $\Im\omega \sim - \vert k\vert ^3c^3/\omega_{\rm p}^2$ in terms of (transverse) wavenumber $k$~\cite{Chang_2008, Lemoine_2015}. In the reference frame of the blast, the shock front moves away, with respect to a given plasma element, at velocity $c/3$ (or $c/2$ is 2D numerical simulations). Hence, damping in time translates into damping in terms of distance to the shock, $x$. 
More specifically, if the one-dimensional power spectrum of the turbulence\footnote{In a 2D simulation, the magnetic turbulence spectrum is defined as $\langle \delta B_z^2 \rangle_y (x) = \int{\rm d}k\,\delta B_k^2$.} satisfies $\langle\delta B_k^2\rangle\,\propto\,k^{-q}$, with $q<1$ (because most of the turbulence power lies on the shortest spatial scales), then the turbulence decays as $\langle\delta B^2(x)\rangle \propto \vert x\omega_{\rm p}/c\vert ^{(q-1)/3}$ for $\vert x\omega_{\rm p}/c\vert\gg1$. Particle-in-cell simulations suggests $\langle\delta B^2\rangle \propto \vert x \vert^{-0.5}$~\cite{Chang_2008,Keshet_2009,Lemoine_2019_PRL}, and therefore a power spectrum index $q\simeq -0.5$, see Fig.~\ref{fig:mudecay} for an illustration.

A decaying microturbulence bears interesting phenomenological consequences for the spectral energy distribution~\cite{Rossi_2003, Derishev_2007, Lemoine_2013a,Lemoine_2015b}. In effect, electrons of Lorentz factor $\gamma$ cool on a synchrotron timescale $t_{\rm syn} \simeq 10^{12}\, \delta B_0^{-2}\gamma^{-1}n_0^{1/2}\,\omega_{\rm p}^{-1}$ (magnetic field $\delta B_0$ expressed in Gauss, density $n_0$ in cm$^{-3}$), thus orders of magnitude larger than $\omega_{\rm p}^{-1}$. All electrons (but those of the very highest energies) thus cool in a region in which the turbulence has decayed through phase mixing. The magnetic field strength that is inferred from the observations, through the modelling of the spectral energy distribution, corresponds to that in the radiation region, and is therefore expected to be much smaller than its effective value in the acceleration region.

To quantify the above effect, one may consider that $\epsilon_B(x)\simeq\epsilon_{B+}$ in a region of width $30-100c/\omega_{\rm p}$ behind the shock front, with $\epsilon_{B+}\simeq 0.01$ the value measured in PIC simulations in the shock vicinity, and that $\epsilon_B$ decays as some powerlaw beyond that distance, $\epsilon_B\propto\left(x \omega_{\rm p}/c\right)^{\alpha}$ (with $\alpha\sim -0.5$), down to a minimal value $\epsilon_{B-}$ near the contact discontinuity.  Incorporating such a model in the computation of gamma-ray burst afterglows indeed produces a satisfactory match to observations for gamma-ray bursts with extended high-energy emission for $\alpha=-0.4$~\cite{Lemoine_2013b}, close to the value seen in PIC simulations.

\begin{figure}
	\begin{center}
\includegraphics[width=0.9\columnwidth]{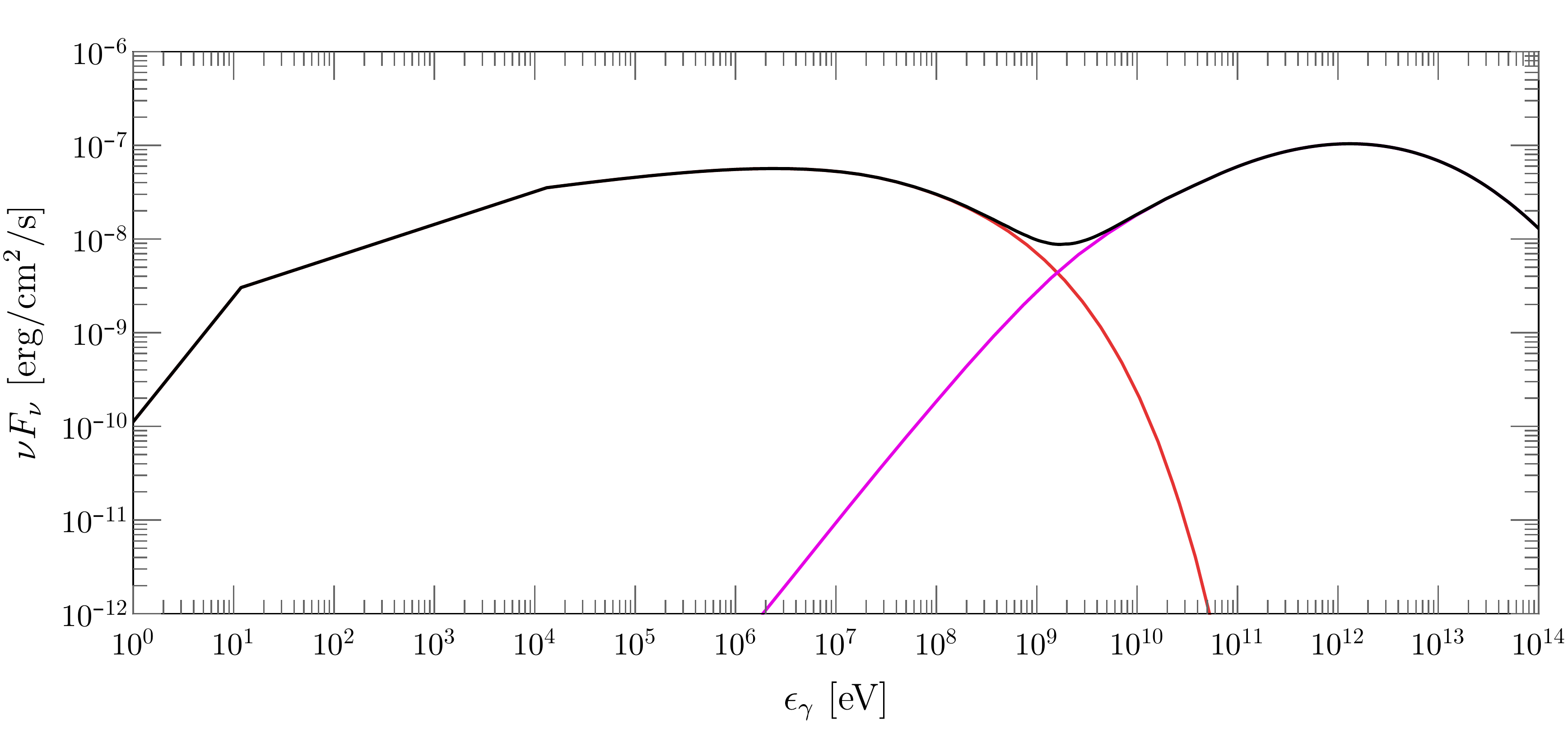}
 \caption{Example of a synchrotron-self-Compton spectrum of a gamma-ray burst (red: synchrotron, magenta: inverse Compton, black: total), at an observer time $t_{\rm obs}=100\,\rm s$ ($t_{\rm obs}=0$ marking the onset of the prompt emission phase), taking into account the effect of a decaying microturbulence behind the shock, as described in the text. The pair-production opacity of the intergalactic radiation fields, which attenuates strongly the emission above $\sim1\,\rm TeV$, has not been taken into account here. See the text for details.}
   \label{fig:mused} 
   	\end{center}
\end{figure}

Another consequence is that those electrons losing their energy through synchrotron on a timescale shorter than the dynamical timescale of the blast, do radiate in a region of changing magnetic field strength. This modifies their synchrotron spectrum and leaves definite signatures in the integrated emission, which could be potentially probed by multiwavelength observational campaigns~\cite{Lemoine_2015b}. Unfortunately, for gamma-ray burst afterglows at least, most of this difference takes place in the hard X-ray - soft gamma-ray regime, which represents the most challenging energy range for instrumentation.

Another generic consequence of the above microphysics is Compton dominance, since a weak magnetic field in the radiation region implies that electrons cool mainly through inverse Compton scattering off the synchrotron-produced photons. As an example, Fig.~\ref{fig:mused} shows the spectral energy distribution of a gamma-ray burst afterglow, at an (observer) timescale of $100\,$s, with $\epsilon_{B+}=0.01$ (value in the shock vicinity) and a decay law $\epsilon_{B}\propto\left(x\omega_{\rm p}/c\right)^{-0.4}$. The other parameters are: energy of the blast wave $E=8\times 10^{53}\,\rm erg$, redshift $z=0.4$, density of the interstellar medium $n=0.03\,\rm cm^{-3}$, electron energy fraction $\epsilon_e=0.1$, accelerated powerlaw index $s=2.3$, and a maximum Lorentz factor $\gamma_{e,\rm max}=2 \times 10^7$, similar to that derived above. The red line presents the synchrotron spectrum, which typically extends up to the GeV range at this early timescale, as discussed above, while the magenta line shows the inverse Compton spectrum.

These parameters have not been chosen at random, but lie very close to those quoted for the modelling of the recent GRB190114C which has been detected up to sub-TeV energies by the MAGIC telescope~\cite{Acciari_2019}, and indeed it is possible to check that the above spectral energy distribution reproduces qualitatively well that observed at early times. Importantly, all of the input microphysical parameters ({i.e.}, $\epsilon_e$, $\epsilon_B$, $s$ and $\gamma_{e,\rm max}$), are based on, or derived from, the physical model described in previous sections. Finally, note that the afterglow model of Ref.~\cite{Acciari_2019} paper assumes a uniform (non-decaying) microturbulence with $\epsilon_B=8\times 10^{-5}$ but, interestingly, our above values $\epsilon_{B+}=0.01$ and $\epsilon_B\propto \left(x\omega_{\rm p}/c\right)^{-0.4}$ lead to $\epsilon_{B-}=6\times 10^{-5}$ in the emission region.

\section{Open questions}

Although the recent decades have known substantial progress in our understanding of the microphysics of relativistic, collisionless shock waves, a number of open issues, with potentially important consequences for phenomenology, remain open. Among those:

\begin{enumerate}
    \item Current understanding suggests that particle acceleration should become efficient at low magnetization, $\sigma \lesssim 10^{-3}$, and indeed this matches well what is seen in PIC simulations. However, the phenomenological modelling of pulsar wind nebulae suggests both that the magnetization at the termination shock is significantly higher than the above critical threshold, and that acceleration is near optimal in those objects, because they are seemingly able to accelerate particles up to the synchrotron burn-off limit (see {\it e.g}~\cite{Amato_2020} and references therein). Observations suggest that particle acceleration takes place close to the termination shock, hence our current theory of relativistic shock acceleration may be currently missing an important item. \\
    
    \item It is important to keep in mind that PIC simulations have so far assumed idealized conditions, meaning a smooth, laminar cold upstream plasma, and that they have been restricted to timescales orders of magnitude below those probed by astrophysical observations. They have also neglected the possible feedback of radiation on the shock structure. The impact of a pre-existing, upstream turbulence, or high amplitude waves, is an important avenue of study for PIC simulations, as this might alter the picture developed earlier. Similarly, any broadband spectrum of radiation generated by the accelerated particles themselves, or by an external source, may have nontrivial and important consequences for the shock physics, through the possible generation of pairs in the shock vicinity (see {\it e.g.}~\cite{Derishev_2016}). \\
    
    \item More generally speaking, how the precursor evolves on long timescales remains a question of debate. The scaling of the microturbulence amplitude in the shock precursor in the PIC simulations depicted above suggests that some form of saturation has been reached, which, in turn, suggests that such precursors could be extended over arbitrarily long length scales, without altering much of their appearance. At the present stage, one cannot exclude that secondary instabilities, {\it e.g.} \cite{Medvedev_2005, Milosavljevic_2006a, Couch_2008, Vanthieghem_2018}, would grow on top of the primary (CFI, PCI) instability responsible for the shock dynamics in the above models. This could modify the turbulence properties in various ways, by changing its strength and/or its coherence length, with direct consequences for acceleration and phenomenology.
      
\end{enumerate}

\section{Summary, final comments}\label{sec:summ}

This paper has provided a status report on the physics of weakly magnetized, relativistic shock waves and their phenomenology, of wide applicability in high-energy and multi-messenger astrophysics.

As the turbulence that sustains the particle acceleration process is induced by the accelerated particles themselves, through electromagnetic microinstabilities driven in the shock precursor, the phenomenology of these shock waves ({\it e.g.}, the multi-messenger signals that they produce) depends strongly on the microphysical processes at play. In this paper, we have thus reviewed in some detail the associated microphysics, in order to extract definite predictions for the `parameters' commonly used in phenomenological studies, namely $\xi_{\rm b}$, the suprathermal electron energy fraction, $\epsilon_B$, the effective magnetization in the shock vicinity, $s$, the suprathermal powerlaw index, as well as the maximal Lorentz factors of the accelerated particles.

Such microphysics is governed by a few essential parameters, namely, the shock four-velocity $u_{\rm sh}$ (relative to the ambient plasma), the  magnetization $\sigma$, and the composition of the ambient (unshocked) plasma. We have separated the discussion according to magnetization, because different instabilities dominate in different domains.

At $\sigma \lesssim 10^{-4}$, the shock physics is dominated by the current filamentation instability (CFI), often referred to as the Weibel instability, which produces a microturbulence on skin-depth scales, elongated into filaments oriented along the direction of shock propagation. At $10^{-4} \lesssim \sigma \lesssim 10^{-2}$, the main instability is a perpendicular current driven instability (PCI) triggered by the transverse electric current, generated by the accelerated particles as they gyrate around the background magnetic field. In both cases, we have sketched a theoretical model of the dynamics of the unshocked plasma. We have shown, in particular, that the shock dynamics becomes independent of the shock Lorentz factor when $\gamma_{\rm sh} \gg 1$, as the precursor, permeated by accelerated particles, plays the role of a buffer that decelerates the background plasma toward a universal trajectory in the shock rest frame.

We have then reviewed a number of phenomenological consequences of direct interest to high-energy astrophysics in Section~\ref{sec:phen}. Some of the most most relevant are: 
\begin{enumerate}
\item The electron distribution exhibits a large minimum Lorentz factor close to $\gamma_{\rm sh}m_p/m_e$, if the shock propagates into a plasma with electrons and protons in equal numbers, due to a nontrivial energy transfer between the ions and the electrons of the unshocked plasma, in the shock precursor; by contrast, in a pair shock, the minimum Lorentz factor is of the order of $\gamma_{\rm sh}$.\\

\item The acceleration rate is significantly slower than Bohm, implying reduced maximal energies for electrons and ions as compared to naive expectations, although large enough, {\it e.g.}, to produce GeV synchrotron (and TeV inverse Compton) photons, or PeV protons during the early afterglow of gamma-ray bursts.\\

\item The small-scale nature of the turbulence makes it prone to decay through phase mixing, which implies that the effective magnetization $\epsilon_B$ may be much smaller in the emission region than in the acceleration region, where it is typically of the order of 0.01.\\ 

\item The ensuing low value of $\epsilon_B$ in the emission region implies that inverse Compton scattering off the synchrotron-produced photons becomes the dominant source of electron cooling, leading to a significant amount of radiation at the highest energies.\\
\end{enumerate}

Most of the recent progress has derived from a fruitful interplay between numerical PIC simulations, theoretical developments and astrophysical inference. As PIC simulations remain currently limited in computing time, in geometry, as well as in their assumptions, one of the great challenges of the coming decade is to transgress these frontiers, and to bridge the gap, in temporal and spatial scales, between the microphysics of kinetic plasma physics and the macrophysics of the source magnetohydrodynamics.

\vspace{6pt} 



\authorcontributions{All authors have contributed equally to this work.}

\funding{We acknowledge financial support from the Programme National Hautes \'Energies (PNHE) of the C.N.R.S.; the ILP Labex (reference ANR-10-LABX-63) as part of the Idex SUPER (reference ANR-11-IDEX-0004-02); the DIWINE Emergence SU 2019 programme;  the ANR-14-CE33-0019 MACH project; and the Plas@Par Labex (reference ANR-11-IDEX-0004-02). This work was supported in part by the National Science Foundation under Grant No.~NSF PHY-1748958. This work was granted access to the HPC resources of TGCC/CCRT under allocations 2018-A0030407666 and 2017-x2016057678 made by GENCI. We also acknowledge PRACE for awarding us access to resource Joliot Curie-SKL at TGCC/CCRT. AV and AG were also partially supported by the U.S. DOE Early Career Research Program under FWP 100331.}

\acknowledgments{}

\reftitle{References}


\externalbibliography{yes}
\bibstyle{mdpi}
\bibliography{Bib}





\end{document}